\documentclass[12pt,thmsa]{article}
\usepackage{amssymb}

\usepackage{amsmath,amssymb}

\input{tcilatex}

\begin{document}

\title{Massive Vector Mesons and Gauge Theory}
\author{Michael D\"{u}tsch\thanks{%
Work supported by the Alexander von Humboldt Foundation and the Deutsche
Forschungsgemeinschaft} \\
II Institut fuer Theoretische Physik\\
Universit\"{a}t Hamburg\\
Luruper Chaussee 149\\
D-22761 Hamburg, Germany\\
email: duetsch@mail.desy.de \and Bert Schroer \\
Institut f\"{u}r Theoretische Physik der FU-Berlin\\
presently: CBPF, Rua Dr. Xavier Sigaud, 150\\
22290-180 Rio de Janeiro, RJ, Brazil\\
email: schroer@cbpfsu1.cat.cbpf.br}
\date{June 1999 }
\maketitle

\begin{abstract}
We show that the requirements of renormalizability and physical consistency
imposed on perturbative interactions of massive vector mesons fix the  
theory essentially uniquely. In particular physical consistency demands the
presence of at least one additional physical degree of freedom which was not
part of the originally required physical particle content. In its simplest
realization (probably the only one) these are scalar fields as envisaged by
Higgs but in the present formulation without the ``symmetry-breaking Higgs
condensate''. The final result agrees precisely with the usual quantization
of a classical gauge theory by means of the Higgs mechanism. But the
emphasis is shifted: instead of invoking the gauge principle (and the Higgs
mechanism) on the local quantum field theory, the principles of local
quantum physics restricted by the perturbative renormalizability demand
``explains'' (via Bohr's correspondence) the classical gauge principle as a
selection principle among the many a priori (semi)classical possibilities of
coupling vector fields among each other. Our method proves an old conjecture
of Cornwall, Levin and Tiktopoulos stating that the renormalization and
consistency requirements of spin=1 particles lead to the gauge theory
structure (i.e. a kind of inverse of 't Hooft's famous renormalizability
proof in quantized gauge theories) which was based on the on-shell unitarity
of the $S$-matrix. Since all known methods of renormalized perturbation
theory are off-shell, our proof is different and a bit more involved than
the original arguments. We also speculate on a possible future ghostfree
formulation which avoids ''field coordinates'' altogether and is expected
to reconcile the on-shell S-matrix point of view with the off-shell field
theory structure.
\end{abstract}

\section{Introduction}

In the development of understanding of the renormalization aspects for
spin=1 interactions, the classical concepts of gauge and fibre bundles have
played a crucial role. Without the closely related Higgs mechanism it would
be hard to imagine, how in the stage of QFT at the end of the 60$^{ies}$,
the incorporation of the electro-weak interaction into the framework of
renormalizable field theory could have been achieved.

The main motivation for the present article starts from the remark that to
our knowledge the Higgs mechanism via Higgs condensates within the setting
of gauge theories has not a well-understood intrinsic quantum physical
content. Hence we choose an alternative procedure by which we will construct
the same physical results for interacting massive vector mesons in a quite
different way which does not rely on the above concepts. In our approach
based on the well-known real-time causal formulation of perturbative QFT,
the renormalizability in the presence of selfinteracting vector mesons is
the basic input requirement and the uniqueness and the gauge appearance of
the polynomial interaction densities are consequences. With other words we
would phrase 't Hooft's famous statement that gauge structure implies
renormalizability the other way around: maintaining perturbative
renormalizability in the presence of vector mesons explains the gauge
structure. Since local quantum theory is more fundamental than classical,
this brings interacting vector mesons into harmony with the properly
understood Bohr's correspondence principle: it is the quantum theory which
tells the classical which possibility among many couplings involving vector
fields and lower spin fields it has to follow, namely the gauge invariant
one.

We would not have gained much, and a cynic might claim that we have replaced
one mystery (the gauge recipe) by another one (the renormalization
prescription), but fortunately, we have some slightly more tangible results
to offer. Our method brings into the open the long looming suspicion that
the appearance of additional physical degrees of freedom (the alias Higgs
particle but now without vacuum condensates) is a necessity, following from
perturbative consistency up to second order (no claim outside of
perturbation theory is made!)\footnote{%
Unfortunately this is not a structural theorem as e.g. the theorem
connecting spontaneous symmetry-breaking with the appearance of ``Goldstone
modes'' or the theorem connecting relating the appearance of ``massive
photons'' with charge-screening; rather it still remains in the setting of
renormalized perturbation theory \cite{Sw}.}. In addition it suggests
strongly that the physics of zero mass theory should be approached from
massive vector mesons, the latter being conceptually (but not necessarily
analytically) simpler. So as it happens often in physics, the new aspect
does not so much lie in the physical results as such, but rather in the
novel way in which they are obtained and in the interpretation associated
with this derivation.

To avoid misunderstandings we emphasize that our approach supports the
principle of local gauge invariance because we obtain the same results
(Lie-algebraic structure, necessity of the scalar ``Higgs'' field) by
starting from quantum physical requirements instead of the conventional,
semiclassical, differential geometric ideas. However it is also important to
emphasize that crucial theoretical observations have been made in the first
5 years of existence of gauge theories (e.g. 't Hooft, Veltmann \cite{tHV}, 
\cite{'t Hooft} and Becchi, Rouet, Stora \cite{B-R-S}). After almost 25
years there seems to be a consensus (based on the present theoretical
aspects of the Standard model) that gauge theory, despite its impressive
early successes, did not (yet) reach the level of maturity of other great
theoretical developments in this century. Our present attempt at the physics
of vector mesons outside (but not in contradiction to) gauge theory is very
much motivated by this state of affairs. We consider this attempt as
something preliminary. Although it starts from the very physical Wigner
particle concept, it shares with the gauge approach the necessity of
extension by unphysical degrees of freedom (ghosts), which in this case
originates from renormalizability. This can be formulated as a cohomological
extension within the Wigner one particle spaces. The advantage of the
present approach is that the auxiliary role of the ghosts as
``renormalization catalyzers''\footnote{%
The ghosts are objects outside local quantum physics which are neither
present in the original physical degrees of freedom (which one wants to
couple) nor in the final physical cohomological subspace. But without their
intermediate presence one does not know how to set up the renormalization
machinery. In the perturbative framework we shall need fermionic and bosonic
(scalar) ghost fields. The first correspond to the ``Faddeev-Popov ghosts'',
the latter to the ``Higgs ghosts'' (which are also called ``would be
Goldstone bosons'') or ``St\"uckelberg fields''.} becomes more manifest. 
The reader finds some speculative remarks on where to look for a ghostfree
formulation in the last sections.

Since some of our concepts are not new, it may be helpful to remind the
reader of some historical precedents. Ever since theories in which vector
mesons or higher spin particles became physically relevant in the late 50$%
^{ies}$, there were two points of views to deal with such problem: either to
start from the Wigner particle picture and stay close to particles and
scattering theory, or to quantize classical gauge field theory (canonically
or by functional integrals) and to make contact with (infra)particles at a
later stage. In fact Sakurai, who introduced the term ``Yang-Mills theory'' 
\cite{Sa}, and who wanted to use quantized classical field theory for the
description of massive vector mesons in strong interactions, encountered
difficulties to reconcile the two points of view. In most of his
contributions he therefore took a phenomenological non-geometric point of
view within a dispersion theoretical setting. This was particularly
advisable since, as a result of some criticism of Pauli, the use of the
Yang-Mills model for the description of massive vector mesons was cast into
question.

The discovery of the electro-weak theory and the renormalization of
Yang-Mills theories \cite{'t Hooft} led to a drastic change of that picture.
The first approach, which tried to make the gauge principle for vector
mesons more palatable from a particle physics point of view, found some
protagonists in the early days of gauge theories (Lewellyn-Smith \cite{Lew},
Bell \cite{B}, Cornwall et al. \cite{C}). It was argued that ``tree
unitarity'' of the $S$-matrix implies that one can choose the fields such
that the model describes a spontaneously broken gauge theory. Cornwall,
Levin and Tiktopoulos even gave a heuristic argument that renormalizability
requires tree unitarity. On the other side the gauge or Yang-Mills point of
view enjoyed general popularity and became the predominant one, irrespective
of whether it could be derived from a more particle dominated on-shell
approach or not. Here we show that the tree unitarity viewpoint may be
substituted by renormalizability within a suitable cohomologically extended
Wigner particle setting. The presence of additional physical degrees of
freedom (Higgs particles) follows from perturbative consistency, even though
there is no concept of gauge symmetry breaking.

The main reason in favor of the gauge point of view was not only its success
in describing problems of quantum matter coupled to external electromagnetic
fields as well as its aesthetical mathematical appeal, but rather that it
opened a path to a renormalizable theory of massive vector mesons, as it was
demonstrated by 't Hooft and Veltman \cite{tHV} \cite{'t Hooft}. This
created, at least for some physicists, the impression that in addition to
causality and spectral properties the general framework of QFT needs an
additional principle for interacting massive vector mesons, namely the gauge
principle. Here we propose a more intrinsic alternative framework which
produces the same physical (gauge invariant) results solely on the
principles of \textbf{l}ocal \textbf{q}uantum \textbf{p}hysics (LQP, for an
introduction and overview see \cite{Ha}) and perturbative renormalizability.
It uses a simplified free form of the BRS formalism \cite{B-R-S} for a
cohomological extension of the Wigner representation theory as a
mathematically less formal operator substitute for the Faddeev-Popov
formalism \cite{F-P} in functional representations. This special
cohomological extension (which is only available for \textit{massive} vector
mesons!) lowers the operator dimension of free massive vector fields from
their physical value 2 to the formal (classical) value 1 without loosing
their point-like nature. Such a lowering is motivated by renormalizability
within the framework of causal perturbation and the resulting theory is
for all practical purposes equivalent to the gauge invariant Lagrangian
approach\footnote{%
Although most of the free fields which correspond to the (m,s) Wigner
representation cannot be obtained in the setting of free Lagrangians $%
\mathcal{L}_{0},$ one can always rewrite a causal interaction $W$ in terms
of Lagrangian field coordinatizations. This is necessary for path integral
formulations, but not for the causal formulation of perturbations.}. Our
contention is that the renormalizability of interacting massive vector
mesons alone uniquely fixes the theory, including its additional Higgs-like
matter content. In other words the gauge and Higgs aspect from which one
starts the Lagrangian quantization in the standard formulation is reduced to
the concept of ``renormalizability'' with which we feel more comfortable
because it occurs on the more fundamental level of local quantum physics
where there is still a future chance for a profound understanding. The
remaining intrinsic feature of the Higgs mechanism is the Schwinger charge
screening\footnote{%
Schwinger, in a little noticed paper and some more extensive published
lecture notes \cite{Schw} thought about a massive phase in QED through the
mechanism of charge screening but without (Higgs) vacuum condensates. In
order to make his non-perturbative ideas of massive vector mesons (the
Maxwellian interaction is renormalizable) more palatable, he invented the
2-dim. Schwinger model. (In fact in the Lowenstein-Swieca treatment of this
model there is a chiral condensate (coming from the $\theta $-degeneracy),
but after the dust has settled, the physical content is described in terms
of a massive free field only). Schwinger's ideas were later elevated by
Swieca into a theorem about vector mass generation via charge screening \cite
{Sw}.} caused by the Higgs-like matter degrees of freedom.

In more recent times a similar point of view has been taken up by Aste,
Scharf and one of the authors (M.D.) \cite{As}. We show however that their
``free perturbative operator gauge invariance'', which is a pure quantum
formulation of gauge invariance, is 
not needed as an a priori perturbative requirement for pure massive
theories: our physical consistency is a weaker requirement but fixes the
theory to the same extent. Some aspects of our point of view appeared in the
work of Grigore \cite{G}. It also has been mentioned in the setting of
Schwinger's work and of general quantum field theory by one of the authors
(B. S) \cite{Schr1}.

Recently there has been an approach to understand the local observable $\ast 
$-algebras of nonabelian gauge theories without addressing the particle
content (which in the LQP framework is anyhow part of the separate more
difficult representation theory associated with states on the algebra) \cite
{D-F}. This approach was specially aimed at the zero mass theories with
infrared problems, because the method does not require the existence of an
adiabatic limit. In that case one cannot use the scattering theory of the
physical particles and the BRS operators cannot be written as bilinear
operators in free fields but they receive interacting contributions in every
order. Therefore one has to face the more difficult problem of a changing
position of the physical cohomology space inside the extended space
depending on the perturbative order. The zero mass limit in our approach is
conceptually and analytically complicated because it is expected to lead to
charge liberation (the opposite of the Schwinger-Swieca charge screening)
and the decoupling of the physical consistency (Higgs) degree of freedom.
Those physical matter fields which will be charged in that limit cannot
maintain their point-like localization, rather one expects that their
localization cannot be better than semi infinite space-like (Mandelstam
string-like). In fact the tight relation between these various phenomena
generates the hope that by controlling those off-shell infrared problems one
may actually get an insight into this (even in perturbation theory)
notoriously difficult localization structure.

For reasons of brevity we exemplify our procedure in a particular class,
namely selfinteracting models of massive vector mesons. In section 2 we
review the causal approach for massive vector mesons and its simplification
as a result of the existence of a natural Fock reference space supplied
by\thinspace scattering theory. Our presentation uses a simplified quadratic
BRS formalism. Physical consistency means that the $S$-matrix constructed on
Fock space (which is the extended space, i.e. it includes ghost states)
induces a well defined (unitary) operator on the physical ``sub''space.

As a justification for the introduction of ghost fields we then describe in
section 3 the apparent clash between renormalizability and the operator
dimension dim$A=2$ of the free vector meson operators in the usual causal
setting and its resolution via cohomological ghost extension, which can be
preempted on the level of the Wigner one-particle spaces.

In the fourth section we formulate physical consistency in an alternative
way which does not rely on scattering theory. The existence of the lowest
dimensional interpolating physical (i.e. BRS-invariant) fields (the ones
which we want to describe in our model as observable particles) does not
only fix the form of these fields, but it also determines the interaction
density (including the necessity of the alias Higgs particle) in agreement
with sect. 2. In particular we find that the physical interpolating fields
are composites in the elementary fields of the extended theory.

So the LSZ-type power series in terms of the\textit{\ physical incoming Fock
space operators} are modified (as we comment in sect. 5). This leads to a
loss of the specific iterative law for the perturbative representation of
the coefficient functions in terms of retarded products (for the fields).
Only the general LSZ-like identification of coefficient functions of local
fields involving multiple commutators of the local field with the incoming
free field (generalized formfactors) remains valid \cite{Schwe}. The reason
for this complication, which prevents the interchange of computations with
the descend to the physical fields, is that the Wick-basis used for writing
the latter in terms of linear combinations of composites (including ghosts)
is not a natural basis for the physical fields (i.e. the fields which
commute with the BRS charge $Q$).

Section 5 also contains some still speculative remarks of where one has to
look for, if one wants to have a ghostfree formulation. In view of the fact
that the previous sections have made clear that ghosts behave in some sense
analogous to catalyzers\footnote{%
One introduces them in order to lower the operator dimension of the
interaction density $W$ to the renormalizable value four. This is achieved
by decreasing the dimension of the vector potential from two to one as a
result of an unphysical cohomological extension of the Wigner one-particle
representation theory. After the calculations have been done in the extended
setting, one eliminates the ghosts by cohomological descend (BRS-invariance).%
} in chemistry, this is not an academic problem but really goes to the root
of understanding of renormalizability for higher spin where the standard
causal approach breaks down. We are led to believe that such an approach
must bypass the Bogoliubov-Shirkov transition operator $S(g)$ and be
on-shell i.e. directly deal with the on-shell $S$-matrix and multi particle
formfactors of physical fields.

We emphasize again that we consider the present formalism as transitory
towards a completely ghostfree formulation of interactions involving higher
spin particles. Although it does not achieve this aim, it creates a critical
Bohr-Heisenberg attitude towards the many unobservable and formal aspects of
the standard gauge formalism.

\section{Consistent perturbative construction of the $S$-matrix for massive
vector fields}

The aim of this section is to construct the
St\"{u}ckelberg-Bogoliubov-Shirkov transition functional $S(\mathbf{g})$,
which is the generating functional for the time ordered products of Wick
polynomials. As most functional quantities this object is not directly
observable but it gives rise to fundamental physical observables as the $S$%
-matrix, formfactors and correlation functions of observable fields. The
notation should not be misread as the $S$-matrix by which we always mean the
scattering operator computed with the LSZ or Haag-Ruelle scattering theory.
Our model is that of selfinteracting massive vector mesons. Our procedure is
related to the one of Scharf, Aste and the first author \cite{As}, \cite{Sch}%
, but similarly to a previous discussion by the other author \cite{Schr2}
and to Grigore \cite{G} as well as older articles as Lewellyn-Smith \cite
{Lew}, we simply rely on physical consistency within the framework of local
quantum physics and do not require such technical tools as ''operator gauge
invariance'' although they tend to simplify calculations.

Using the St\"{u}ckelberg-Bogoliubov-Shirkov-Epstein-Glaser method \cite{B-S}
\cite{E-G} we make the following perturbative Ansatz for $S(\mathbf{g})$ 
\begin{equation}
S(\mathbf{g})=\mathbf{1}+\sum_{n=1}^{\infty }\frac{i^{n}}{n!}\int
d^{4}x_{1}...d^{4}x_{n}%
\,T_{j_{1}...j_{n}}(x_{1},...,x_{n})g_{j_{1}}(x_{1})...g_{j_{n}}(x_{n}),%
\quad \mathbf{g}=(g_{j})_{j=0}^{G},  \label{2.1}
\end{equation}
$g_{j}\in \mathcal{S}(\mathbf{R}^{4})$, which is a formal power series in $%
\mathbf{g}$. The unknown $T_{j_{1}...j_{n}}$ are operator valued
distributions \footnote{%
For questions concerning domains we refer to \cite{E-G}. For all operators
which appear in sections 2 and 3 there exists a common dense invariant
domain $\mathcal{D}$ and we restrict all operators to this subspace.}. They
are constructed inductively by means of the following requirements (strongly
influenced by the general Wightman-framework \cite{St-W}):

(A) \textit{Specification of the model in first order}: The first order
expressions are the main input of the construction. They specify the model
and must be local: $[T_{j}(x),T_{k}(y)]=0$ for $(x-y)^{2}<0$. We assume that 
$T_{0}(x)\equiv W_0(x)\equiv W(x)=W(x)^*$ is the physically relevant
interaction density in Fock space (i.e. a Poincar\'{e} covariant scalar
composite described by a Wick polynomial). $T_{j}\equiv W_{j},\,j=1,...,G$
are auxiliary interactions. The interaction $W_{j}$ is switched by the
space-time dependent coupling ''constant'' $g_{j}\in \mathcal{S}(\mathbf{R}%
^{4})$. The physically relevant $S$-matrix is obtained in the adiabatic
limit: $g_{0}\rightarrow \mathrm{const.},\> g_{j}\rightarrow 0,\>j=1,...,G.$

(B) \textit{Permutation symmetry}: Due to the Ansatz (\ref{2.1}) we may
require permutation symmetry 
\begin{equation}
T_{j_{\pi 1}...j_{\pi n}}(x_{\pi 1},...,x_{\pi
n})=T_{j_{1}...j_{n}}(x_{1},...,x_{n}),\quad \quad \forall \pi \in \mathcal{S%
}_{n}.  \label{2.2}
\end{equation}

(C) \textit{Causality}: 
\begin{equation}
S(\mathbf{g}^{(1)}+\mathbf{g}^{(2)})=S(\mathbf{g}^{(1)})S(\mathbf{g}%
^{(2)})\quad \quad \mathrm{if}\quad \quad \cup _{j}\>\mathrm{supp}%
\>g_{j}^{(1)}\>\cap \>(\cup _{j}\>\mathrm{supp}\>g_{j}^{(2)}+\bar{V}%
_{-})=\emptyset ,  \label{2.3}
\end{equation}
where $\bar V_{\mp}$ denotes the closed backward/forward light cone. This
requirement is equivalent to (see the appendix of \cite{D1}) \footnote{%
The non-trivial part of this equivalence is that in the $n$-th order
expression of (\ref{2.3}) only special test functions appear, whereas (\ref
{2.4}) holds on $\mathcal{S}(\mathbf{R}^{4n})$.} 
\begin{equation}
T_{j_{1}...j_{n}}(x_{1},...,x_{n})=T_{j_{1}...j_{l}}(x_{1},...,x_{l})T_{j_{l+1}...j_{n}}(x_{l+1},...,x_{n})
\label{2.4}
\end{equation}
if 
\[
\{x_{1},...,x_{l}\}\cap (\{x_{l+1},...,x_{n}\}+\bar{V}_{-})=\emptyset . 
\]
This means that $T_{j_{1}...j_{n}}(x_{1},...,x_{n})$ is a (well-defined)
time ordered product of $W_{j_{1}}(x_{1}),\break ...,W_{j_{n}}(x_{n})$.
Hence we use the notation 
\begin{equation}
T_{j_{1}...j_{n}}(x_{1},...,x_{n})=T(W_{j_{1}}(x_{1})...W_{j_{n}}(x_{n}))
\label{2.5}
\end{equation}
Due to the induction with respect to the order $n$, the $T_{j_{1}...j_{n}}$
are uniquely fixed by causality up to the total diagonal $\Delta
_{n}=\{(x_{1},...,x_{n})|x_{1}=x_{2}=...=x_{n}\}$. The extension of the $%
T_{j_{1}...j_{n}}$ to the total diagonal is non-unique. It is restricted by
the following normalization conditions:

(D) \textit{Poincar\'{e} covariance};

(E) \textit{Unitarity}: $S(\mathbf{g})^{-1}=S(\mathbf{g})^*$ for $\mathbf{g}%
=(g_0,0,...,0)$, $g_0$ real valued;

(F) \textit{Scaling degree}: The degree of the singularity at the diagonal,
measured in terms of Steinmann' scaling degree \cite{Ste}\cite{Br-F}%
\footnote{%
We adopt here the notion 'scaling degree' to operator valued distributions
by using the strong operator topology. Note that the scaling degree of a
Wick monomial agrees with its mass dimension.}, may not be increased by the
extension. This ensures \textit{renormalizability by power counting} if the
scaling degree (or 'mass dimension') of all $W_{j}$ is $\leq 4$. This degree
is a tool which is related to Weinberg's power counting.

Additional normalization conditions must be imposed, if one wants to
maintain further symmetries or relations \footnote{%
We consider symmetries and relations which are satisfied away from the total
diagonal, due to the causal factorization (\ref{2.5}) and the inductive
assumption. Poincar\'{e} covariance (D) and unitarity (E) are of this type.}%
, e.g. discrete symmetries (P,C,T), 'operator gauge invariance' (\ref{2.29}-%
\ref{2.30}) or the field equations of the interacting fields (\textbf{(N4)}
in \cite{D-F}), which can be obtained from the functional $S(\mathbf{g})$ (%
\ref{2.1}) by Bogoliubovs formula (see sect.4).

The existence of the adiabatic limit restricts the extension additionally:
for \textit{pure massive} theories Epstein and Glaser \cite{E-G1} proved
that, with correct mass and wave function (re)normalization the adiabatic
limit of the functional $S(\mathbf{g})$ exists in the strong operator sense
and it is this limit which we call '$S$-matrix'. More precisely setting $%
\mathbf{g}_{\epsilon }(x):=(g_{0}(\epsilon x),0,...0)$ the limit 
\begin{equation}
S_{n}\psi \equiv \lim_{\epsilon \rightarrow 0}S_{n}(\mathbf{g}_{\epsilon
})\psi  \label{2.6}
\end{equation}
exists $\forall \psi \in \mathcal{D}$, where $g_{0}\in \mathcal{S}(\mathbf{R}%
^{4}),\>g:=g_{0}(0)>0$ is the coupling constant and $S_n(\mathbf{g})$ ($%
S_{n} $ resp.) denotes the $n$-th order of the functional $S(\mathbf{g})$ ($%
S $-matrix resp.). It follows that the $S$-matrix is unitary as an operator
valued formal power series in Fock space \cite{E-G1}: $S=\sum_{n}S_{n},%
\>S_{n}\sim g^{n}$, $S^{\ast }S=\mathbf{1}=SS^{\ast }$ on $\mathcal{D}$.

Due to this fact we solely consider models in which all fields are massive.
In order that it makes physically sense to consider the $S$-matrix, we
assume that there are no unstable physical particles as e.g. the W- and
Z-bosons in the electroweak theory. \vskip0.5cm

In \textit{vector meson theories} (which are ``gauge theories'') the crucial
problem is the elimination of the unphysical degrees of freedom. In the $S$%
-matrix framework this problem turns into the requirement that the $S$%
-matrix induces a well-defined unitary operator on the space of physical
states (this is discussed in detail below). We will see that this condition
is very restrictive: it determines the possible interactions to a large
extent.

Let us first consider the free incoming fields. We quantize the free vector
fields $(A_{a}^{\mu })_{{a}=1,...,M}$ by 
\begin{equation}
(\square +m_{a}^{2})A_{a}^{\mu }=0,\quad \lbrack A_{a}^{\mu }(x),A_{b}^{\nu
}(y)]=ig^{\mu \nu }\delta _{ab}\Delta _{m_{a}}(x-y),\quad A_{a}^{\mu \,\ast
}=A_{a}^{\mu }  \label{2.7}
\end{equation}
(where $\Delta _{m}$ is the Pauli-Jordan distribution to the mass $m$),
which corresponds to the ``Feynman gauge''. The representation of this $*$%
-algebra requires an indefinite inner product space. We, therefore, work in
a Krein Fock space $\mathcal{F}$. We denote the scalar product by $(.,.)$
and $A^{+}$ is the adjoint of $A$ w.r.t. $(.,.)$. Let $J$ be the Krein
operator: $J^{2}=1,\>J^{+}=J$. Then the indefinite inner product $<.,.>$ is
defined by 
\begin{equation}
<a,b>\equiv (a,Jb),\quad \quad a,b\in \mathcal{F}  \label{2.8}
\end{equation}
and $\ast $ denotes the adjoint with respect to $<.,.>$: 
\begin{equation}
O^{\ast }\equiv JO^{+}J,\quad \quad <Oa,b>=<a,O^{\ast }b>  \label{2.9}
\end{equation}
Let $Q$ be an (unbounded) $*$-symmetrical nilpotent operator in $\mathcal{F}$
\begin{equation}
Q=Q^{\ast }\quad (\mathrm{on\,\>the\>\,dense\,\>invariant\>\,domain}\>\,%
\mathcal{D}),\quad \quad Q^{2}=0  \label{2.10}
\end{equation}
By means of $Q^{2}=0$ one easily finds that $\mathcal{D}$ is the direct sum
of three, pairwise orthogonal (w.r.t. $(.,.)$) subspaces \cite{HV}\cite{K} 
\begin{equation}
\mathcal{D}=\mathrm{ran}\>Q\oplus (\mathrm{ker}\>Q\cap \mathrm{ker}%
\>Q^{+})\oplus \mathrm{ran}\>Q^{+}  \label{2.11}
\end{equation}
\begin{equation}
\mathrm{ker}\>Q=\mathrm{ran}\>Q\oplus (\mathrm{ker}\>Q\cap \mathrm{ker}%
\>Q^{+}),\quad \quad \mathrm{ker}\>Q^{+}=\mathrm{ran}\>Q^{+}\oplus (\mathrm{%
ker}\>Q\cap \mathrm{ker}\>Q^{+})  \label{2.12}
\end{equation}
In addition we assume 
\begin{equation}
J|_{\mathrm{ker}\>Q\cap \mathrm{ker}\>Q^{+}}=\mathbf{1}\quad \quad \quad (%
\mathrm{positivity\,\ \ \>\>assumption})  \label{2.13}
\end{equation}
Then the $<.,.>$-product is positive definite on 
\begin{equation}
\mathcal{H}_{\mathrm{phys}}\equiv \mathrm{ker}\>Q\cap \mathrm{ker}\>Q^{+}
\label{2.14}
\end{equation}
and $\mathcal{H}_{\mathrm{phys}}$ is interpreted as the \textit{physical
subspace} of $\mathcal{F}$. We denote the projectors on $\mathrm{ran}\>Q$ ($%
\mathcal{H}_{\mathrm{phys}},\>\mathrm{ran}\>Q^{+}$ resp.) by $P_{-}$ ($%
P_{0},\>P_{+}$ resp.) 
\begin{equation}
\mathbf{1}=P_{-}+P_{0}+P_{+}\quad \quad (\mathrm{on\,\ }\>\mathcal{D})
\label{2.15}
\end{equation}
Note $Q=P_{-}QP_{+}$. The positivity (\ref{2.13}) and $Q=Q^{\ast }$ imply 
\cite{K} 
\begin{equation}
P_{0}=P_{0}JP_{0},\quad \quad J=P_{0}JP_{0}+P_{-}JP_{+}+P_{+}JP_{-}
\label{2.16}
\end{equation}
and hence 
\begin{equation}
P_{0}^{\ast }=P_{0},\quad \quad P_{-}^{\ast }=P_{+}  \label{2.17}
\end{equation}
Let $S:\mathcal{D}\rightarrow \mathcal{D}$ be the (strong) adiabatic limit (%
\ref{2.6}) of $S(g)$. We define 
\begin{equation}
S_{ab}\equiv P_{a}SP_{b},\quad \quad \quad a,b\in \{-,0,+\}  \label{2.18}
\end{equation}
and obtain 
\begin{equation}
P_{a}S^{\ast }P_{b}=(S_{(-b)\,(-a)})^{\ast }  \label{2.19}
\end{equation}
For pedagogical reasons we introduce the matrix notation according to the
decomposition (\ref{2.11}) 
\begin{equation}
J= 
\begin{pmatrix}
0 & 0 & P_{-}JP_{+} \\ 
0 & P_{0} & 0 \\ 
P_{+}JP_{-} & 0 & 0
\end{pmatrix}
\label{2.20}
\end{equation}
\begin{equation}
S= 
\begin{pmatrix}
S_{--} & S_{-0} & S_{-+} \\ 
S_{0-} & S_{00} & S_{0+} \\ 
S_{+-} & S_{+0} & S_{++}
\end{pmatrix}
\label{2.21}
\end{equation}
and 
\[
S^{\ast }= 
\begin{pmatrix}
(S_{++})^{\ast } & (S_{0+})^{\ast } & (S_{-+})^{\ast } \\ 
(S_{+0})^{\ast } & (S_{00})^{\ast } & (S_{-0})^{\ast } \\ 
(S_{+-})^{\ast } & (S_{0-})^{\ast } & (S_{--})^{\ast }
\end{pmatrix}
\]
by means of (\ref{2.19}).

An alternative definition of the physical states is 
\begin{equation}
\mathcal{H}_{\mathrm{phys}}^{\prime }\equiv \frac{\mathrm{ker}\,Q}{\mathrm{%
ran}\,Q}  \label{2.22}
\end{equation}
where the scalar product in $\mathcal{H}_{\mathrm{phys}}^{\prime }$ is
defined such that the map 
\begin{equation}
\mathcal{H}_{\mathrm{phys}}\rightarrow \mathcal{H}_{\mathrm{phys}}^{\prime
}:\phi \rightarrow \lbrack \phi ]  \label{2.23}
\end{equation}
is a pre Hilbert space isomorphism. ($[\phi ]$ denotes the equivalence class
of $\phi $.) The first definition (i.e. $\mathcal{H}_{\mathrm{phys}}$) has
the advantage that the set of physical states is a \textit{sub}space of the
Krein Fock space $\mathcal{F}$, which has a clear particle interpretation.
But $\mathcal{H}_{\mathrm{phys}}$ is not Lorentz invariant (in contrast to $%
\mathcal{H}_{\mathrm{phys}}^{\prime}$). The change of its position inside
the total space under Lorentz-transformations is of course a result of the
lack of Lorentz-invariance of J.

To describe \textit{free} spin=1 fields the introduction of a BRST-formalism
(as we just have done) is not necessary. However, our main topic is to
describe the interacting theory in the adiabatic limit. So the incoming and
outgoing fields are asymptotically free. Hence, being equipped with a
characterization of the asymptotical physical states in terms of a
BRST-formalism for free fields, our scattering point of view is much
simplified.

Let $S^*S=\mathbf{1}=SS^*$ (on $\mathcal{D}$). We now discuss two different
formulations of the physical consistency of the $S$-matrix:

(i) A consistent $S$-matrix theory requires 
\[
P_0S^*P_0SP_0=P_0=P_0SP_0S^*P_0\quad\Longleftrightarrow \quad
(S_{00})^*=(S_{00})^{-1}\quad \mathrm{on}\quad \mathcal{H}_{\mathrm{phys}}. 
\]

(ii) In the framework of the definition (\ref{2.22}) of the physical states
consistency means that $S$ and $S^{-1}=S^{\ast }$ induce well-defined
operators on the factor space $\mathcal{H}_{\mathrm{phys}}^{\prime }$ by the
definition 
\begin{equation}
\lbrack \mathcal{O}][\phi ]\equiv \lbrack \mathcal{O}\phi ],\quad \quad 
\mathcal{O}=S,S^{\ast }  \label{2.24}
\end{equation}
This holds true iff 
\begin{equation}
\mathcal{O}\>\mathrm{ker}\,Q\subset \mathrm{ker}\,Q\quad \quad \wedge \quad
\quad \mathcal{O}\>\mathrm{ran}\,Q\subset \mathrm{ran}\,Q,\quad \quad 
\mathcal{O}=S,S^{\ast }  \label{2.25}
\end{equation}
Due to $[\mathcal{O}]^{\ast }=[\mathcal{O}^{\ast }]$ the physical $S$-matrix 
$[S]$ is then unitary.

The following Lemma states that (i) is a truly weaker condition than (ii)
and gives equivalent formulations of (ii).

\textbf{Lemma 1:} The following statements (a)-(g) are equivalent and they
imply (h). But (h) does not imply the other statements if $Q\not{= }0$.

(a) $SS^*=\mathbf{1}=S^*S$ and $S\mathrm{ker}\,Q\subset \mathrm{ker}\,Q$
(i.e. $S_{+-}=0=S_{+0}$).

(b) $SS^*=\mathbf{1}=S^*S$ and $[Q,S]\vert_{\mathrm{ker}\,Q}=0$.

(c) $SS^*=\mathbf{1}=S^*S$ and $S\mathrm{ran}\,Q\subset \mathrm{ran}\,Q$
(i.e. $S_{+-}=0=S_{0-}$).

(d) The matrix $S$ (\ref{2.21}) has the form 
\begin{equation}
S= 
\begin{pmatrix}
S_{--} & S_{-0} & S_{-+} \\ 
0 & S_{00} & -S_{00}(S_{-0})^{\ast }(S_{--})^{\ast \,-1} \\ 
0 & 0 & (S_{--})^{\ast \,-1}
\end{pmatrix}
\label{2.26}
\end{equation}
where $S_{--}$ and $S_{00}$ are invertible (on $\mathrm{ran}\,Q$, $\mathcal{H%
}_{\mathrm{phys}}$ resp.) and $S_{00},\,S_{--},\,S_{-0},\,S_{-+}$ satisfy 
\[
(S_{00})^{\ast }=(S_{00})^{-1},\quad \quad S_{-+}(S_{--})^{\ast
}+S_{-0}(S_{-0})^{\ast }+S_{--}(S_{-+})^{\ast }=0. 
\]

(e) $SS^*=\mathbf{1}=S^*S$ and $S^*\mathrm{ker}\,Q\subset \mathrm{ker}\,Q$
(i.e. $(S_{+-})^*=0=(S_{0-})^*$).

(f) $SS^*=\mathbf{1}=S^*S$ and $[Q,S^*]\vert_{\mathrm{ker}\,Q}=0$.

(g) $SS^*=\mathbf{1}=S^*S$ and $S^*\mathrm{ran}\,Q\subset \mathrm{ran}\,Q$
(i.e. $(S_{+-})^*=0=(S_{+0})^*$).

(h) $SS^*=\mathbf{1}=S^*S$ and $(S_{00})^*S_{00}=P_0=S_{00}(S_{00})^*$.

\textit{Proof:} (b) $\Leftrightarrow$ (a) $\Leftrightarrow$ (g) and (f) $%
\Leftrightarrow$ (e) $\Leftrightarrow$ (c) hold trivially true.

(a) $\Leftrightarrow$ (e): (a) implies $S_k\mathrm{ker}\,Q\subset \mathrm{ker%
}\,Q$ for each order $S_k$ of $S$. Therefore, 
\[
(S^*)_n=(S^{-1})_n=\sum_{r=1}^n(-1)^r\sum_{n_1,...,n_r\geq
1,\>n_1+...+n_r=n} S_{n_1}...S_{n_r} 
\]
maps $\mathrm{ker}\,Q$ in $\mathrm{ker}\,Q$ and, hence, this holds also true
for $S^*=\sum_n(S^*)_n$. (e) $\Rightarrow$ (a) follows analogously.

(a),(c) $\Leftrightarrow $ (d): By a straightforward calculation one
verifies that the equations 
\begin{equation}
S_{+-}=0,\>S_{+0}=0,\>S_{0-}=0\quad \mathrm{and}\quad
\sum_{c}(S_{(-c)\,(-a)})^{\ast }S_{cb}=\delta
_{ab}P_{a}=\sum_{c}S_{ac}(S_{(-b)\,(-c)})^{\ast }  \label{2.27}
\end{equation}
are equivalent to (d).

(a),(c) $\Rightarrow$ (h): Choosing $a=0=b$ in (\ref{2.27}) we obtain $%
(S_{00})^*S_{00}=P_0=S_{00}(S_{00})^*$.

To show that (h) does not imply the other statements for $Q\not{= }0$ ($%
\Leftrightarrow \mathrm{ran}\,Q\not{= }0\Leftrightarrow \mathrm{ran}\,Q^+%
\not{= }0$), we give two examples for $S$ which satisfy (h) but not (d):

- The condition (h) is invariant under an exchange of $Q$ and $Q^+$.
Therefore, there exists a solution S of (h) which maps $\mathrm{ker}\,Q^+$
in $\mathrm{ker}\,Q^+$ (and/or $\mathrm{ran}\,Q^+$ in $\mathrm{ran}\,Q^+$),
i.e. $S$ is a lower triangular matrix.

- The following $S$-matrix fulfills (h) and $S_{+-}\not{=}0\not{=}S_{-+}$
(if $cd\not{=}0$): 
\begin{equation}
S= 
\begin{pmatrix}
ae^{i\alpha }P_{-} & 0 & ice^{i\alpha }P_{-}JP_{+} \\ 
0 & e^{i\phi }P_{0} & 0 \\ 
ide^{i\alpha }P_{+}JP_{-} & 0 & be^{i\alpha }P_{+}
\end{pmatrix}
\label{2.28}
\end{equation}
with $a,b,c,d\in \mathbf{R},\>\>\alpha ,\phi \in \mathbf{R}$ and $ab+cd=1$. $%
\quad \square $

In references \cite{As}, \cite{Sch} physical consistency of the $S$-matrix
is satisfied by requiring a perturbative condition which implies $[Q,S]=0$,
namely the \vskip0.2cm \textbf{'free perturbative operator gauge invariance'}
\footnote{%
The reason for this name stems from the fact that $Q$ is the generator of
the BRST-transformation of the free incoming fields. For the present purpose
we will simply refer to it as the ``Q-divergence condition'' for time
ordered products since it has nothing to do with the classical notion of
gauge invariance in the differential geometric setting of fibre bundles, but
secures that the true $S$-matrix is physical. Roughly speaking it is the off
shell version of physicality of the on-shell $S$-matrix.} (or \textbf{'$Q$%
-divergence condition'}): Let $W\equiv T_{0} $ be the interaction
Lagrangian. Then there exists a Wick polynomial $W_{1}^{\nu }$ with 
\begin{equation}
\lbrack Q,W(x)]=i\partial _{\nu }W_{1}^{\nu }(x)  \label{2.29}
\end{equation}
and the time ordered products of $W$ and $W_{1}^{\nu }$ fulfil 
\begin{equation}
\lbrack Q,T(W(x_{1})...W(x_{n}))]=i\sum_{l=1}^{n}\partial _{\mu
}^{x_{l}}T(W(x_{1})...W_{1}^{\mu }(x_{l})...W(x_{n}))  \label{2.30}
\end{equation}
\vskip0.2cm Let us assume that (\ref{2.30}) holds true to all orders $\leq
(n-1)$. Due to the causal factorization (\ref{2.4}) the requirement (\ref
{2.30}) is then satisfied away from the total diagonal, i.e. on $\mathcal{S}(%
\mathbf{R}^{4n}\setminus \Delta _{n})$. Hence, (\ref{2.30}) is an additional
normalization condition for $T(W...W)$ and $T(W...W_{1}^{\mu }...W)$. But it
is a highly non-trivial task to prove that there exists an extension to the
diagonal which satisfies (\ref{2.30}) and the other normalization conditions
(D), (E) and (F).

In contrast to $[Q,S]=0$ or $[Q,S]\vert_{\mathrm{ker}\,Q}=0$, the $Q$%
-divergence condition (\ref{2.29}-\ref{2.30}) makes sense also in models in
which the adiabatic limit does not exist, e.g. for massless selfinteracting
vector fields (``nonabelian gauge theories''). For massless $SU(N)$%
-Yang-Mills theories it has been proved that the $Q$-divergence condition (%
\ref{2.29}-{\ref{2.30}) (more precisely the corresponding C-number
identities which imply (\ref{2.29}-\ref{2.30})) can be satisfied to all
orders \cite{DHS},\cite{D2}, and that these C-number identities imply the
usual Slavnov-Taylor identities \cite{D3}. In addition, (\ref{2.29}-\ref
{2.30}) determines to a large extent the possible structure of the model
(see below). We emphasize that this is a \textit{pure quantum formulation of
gauge invariance}, without reference to classical physics. }

In the present case of pure massive vector mesons (or more generally in
``gauge'' theories in which the strong adiabatic limit (\ref{2.6}) of the
transition functional $S(\mathbf{g})$ exists) we proceed in an alternative
way. \textit{We do not require the $Q$-divergence condition (\ref{2.29}-\ref
{2.30}) as a new physical principle, instead we simply require physical
consistency of the $S$-matrix}, which means 
\begin{equation}
\lbrack Q,S]\vert_{\mathrm{ker}\,Q}=0  \label{2.31}
\end{equation}
by Lemma 1. This is a weaker condition than (\ref{2.29}-\ref{2.30}). But we
will see that it determines the theory to the same extent. Our procedure is
similar to Grigore \cite{G}.\footnote{%
However, there are two conceptual shortcomings in these papers. It is
overlooked that most of the trilinear terms in $W$ vanish in the adiabatic
limit due to energy-momentum conservation and, hence, to first order the
condition (\ref{2.31}) yields no information about the trilinear terms in $W$
(see appendix A).
\par
By using the terminology introduced below the second shortcoming can be
described as follows: the Higgs field(s) is/are treated as scalar partner(s)
(with arbitrary mass $m_{H}\geq 0$) of the massless vector field(s), which
does/do not appear in $Q$ and, hence, is/are physical. By chance this works
for the electroweak theory (there is one massless vector field and one Higgs
field is needed). But e.g. in the present case of pure massive vector
mesons, there would be no Higgs field and such a model is physically
inconsistent (\ref{2.31}) to second order. This insufficiency does not
appear in \cite{As} and \cite{Sch}.}

First we construct the operator $Q$ (\ref{2.10}) which defines the physical
states. For massless vector meson theories the procedure is well-known \cite
{K-O}, \cite{DHKS}. The nilpotency of $Q$ gives reason to introduce an
anticommuting pair of ghost fields $u_{a},\tilde{u}_{a}$ for each vector
field $A_{a}$ (fermionic ghosts). Then we define\footnote{%
The convergence of this integral (and also of the corresponding expression (%
\ref{2.36}) in the massive theory) can be shown by using a method of
Requardt \cite{R},\cite{D-F}.} 
\begin{equation}
Q\equiv \int d^{3}x\,\sum_{a}\partial _{\nu }A_{a}^{\nu }(x)%
\overleftrightarrow{\partial }_{0}u_{a}(x).  \label{2.32}
\end{equation}

Turning to massive vector fields $A_{a}$ (\ref{2.7}) we give the fermionic
ghost fields the same masses $m_{a}$ (otherwise the current $%
\sum_{a}\partial _{\nu } A_{a}^{\nu }\overleftrightarrow{\partial }%
_{\mu}u_{a}$ or $\sum_{a}(\partial _{\nu }A_{a}^{\nu }+m_{a}\phi_{a}) 
\overleftrightarrow{\partial }_{\mu}u_{a}$ (see below) would not be
conserved) 
\[
(\square +m_{a}^{2})u_{a}=0,\quad (\square +m_{a}^{2})\tilde{u}_{a}=0,\quad
\{u_{a}(x),u_{b}(y)\}=0,\quad \{\tilde{u}_{a}(x),\tilde{u}_{b}(y)\}=0, 
\]
\begin{equation}
\{u_{a}(x),\tilde{u}_{b}(y)\}=-i\delta _{ab}\Delta _{m_{a}}(x-y),\quad
u_{a}^{\ast }=u_{a},\quad \tilde{u}_{a}^{\ast }=-\tilde{u}_{a}.  \label{2.33}
\end{equation}
If we insert the massive $A_{a}$ and $u_{a}$ fields into the formula (\ref
{2.32}) for $Q$ the nilpotency is lost 
\begin{equation}
2Q_{\mathrm{naive}}^{2}=\{Q_{\mathrm{naive}},Q_{\mathrm{naive}}\}=\int
d^{3}x\,\int d^{3}y\,[\partial _{\mu }A_{a}^{\mu }(x),\partial _{\nu
}A_{b}^{\nu }(y)]\overleftrightarrow{\partial }_{x_{0}}\overleftrightarrow{%
\partial }_{y^{0}}u_{a}(x)u_{b}(y)\not{=}0.  \label{2.34}
\end{equation}
To restore the nilpotency we proceed as follows \cite{K}: to each vector
field $A_{a}$ we consider a scalar field $\phi _{a}$ with the same mass $%
m_{a}$ 
\begin{equation}
(\square +m_{a}^{2})\phi _{a}=0,\quad \lbrack \phi _{a}(x),\phi
_{b}(y)]=-i\delta _{ab}\Delta _{m_{a}}(x-y),\quad \phi _{a}^{\ast }=\phi
_{a}.  \label{2.35}
\end{equation}
We call the fields $\phi_a$ also 'ghost fields' (bosonic ghosts) because
they are unphysical (see below). Then, due to $[\partial _{\mu }A_{a}^{\mu
}+m_{a}\phi _{a},\partial _{\nu }A_{b}^{\nu }(y)+m_{b}\phi _{b}]=0$, the
charge 
\begin{equation}
Q\equiv \int d^{3}x\,\sum_{a}(\partial _{\nu }A_{a}^{\nu }(x)+m_{a}\phi
_{a}(x))\overleftrightarrow{\partial }_{0}u_{a}(x)  \label{2.36}
\end{equation}
is nilpotent and symmetrical. Later we shall see that an additional scalar
field $H$ with arbitrary mass $m_{H}\geq 0$ is needed\footnote{%
For simplicity we only consider models in which \textit{one} additional
scalar field suffices for a consistent construction of the $S$-matrix.} 
\begin{equation}
(\square +m_{H}^{2})H=0,\quad \lbrack H(x),H(y)]=-i\Delta
_{m_{H}}(x-y),\quad H^{\ast }=H.  \label{2.37}
\end{equation}
The notation $H$ is reminiscent of the \textit{Higgs} field, but one with no
``vacuum condensate'' (leaving aside the academic point of whether Higgs
idea would allow for our terminology): $<0|H(x)|0>=0$, where $|0>\in 
\mathcal{F}$ is the vacuum of the free fields. For the representation of $%
A_{a}$ (\ref{2.7}), $u_{a},\tilde{u}_{a}$ (\ref{2.33}), $\phi _{a}$ (\ref
{2.35}) and $H$ (\ref{2.37}) in a Krein Fock space $\mathcal{F}$ and
especially the definition of $J$ we refer to \cite{K}\footnote{%
In \cite{K} it was not realized that without the Higgs field $H$ (or an
ingenious substitute) the $Q$-divergence condition to second order is
violated. We represent the $H$-field similarly to the other scalar fields $%
\phi _{a}$ with $J=\mathbf{1}$ in the $H$-Fock space.}. Note that the ghost
fields $u_a,\tilde u_a$ (\ref{2.33}) and $\phi_a$ (\ref{2.35}) are also
asymptotic fields, but that asymptotic states containing ghost exitations do
not belong to the physical Hilbert space (\ref{2.14}). From the commutation
relations 
\[
\lbrack Q,A_{a}^{\mu }]=i\partial ^{\mu}u_{a},\quad \lbrack Q,\phi
_{a}]=im_{a}u_{a},\quad \{Q,u_{a}\}=0, 
\]
\begin{equation}
\{Q,\tilde{u}_{a}\}=-i(\partial _{\mu }A_{a}^{\mu }+m_{a}\phi _{a}),\quad
\lbrack Q,H]=0  \label{2.38}
\end{equation}
we conclude that $\mathcal{H}_{\mathrm{phys}}$ (\ref{2.14}) is the linear
span of the set of states $B_{1}....B_{l}|0>,\>l\in \mathbf{N}_{0}$, where $%
B_{1},....,B_{l}$ are transversal vector meson fields (three polarizations)
or $H$-fields. Using this explicit result the positivity assumption (\ref
{2.13}) can be verified \cite{K}. We emphasize that $H$ is physical, in
contrast to the (scalar) bosonic ghost fields $\phi _{a}$.

We are now looking for the possible interactions $W\equiv W_0$ which satisfy
the following requirements:

(a) $W$ is a Wick polynomial in the free incoming fields. Each monomial in $%
W $ has at least three factors,

(b) $W$ is invariant with respect to Poincar\'{e} transformations,

(c) the number of $u$-fields agrees with the number of $\tilde{u}$-fields in
each monomial of $W$ (i.e. the 'ghost number' is zero),

(d) the scaling degree (or mass dimension) of $W$ is $\leq 4$ (this is
necessary for renormalizability by power counting),

(e) $W=W^*$ (which yields $S^*=S^{-1}$ if the time ordered products are
suitably normalized),

(f) physical consistency (\ref{2.31}).

First we point out that the requirements (a)-(f) do not fix the interaction
uniquely. Even if we replace in (f) the physical consistency (\ref{2.31}) by
the $Q$-divergence condition (\ref{2.29}-\ref{2.30}), which is a stronger
requirement, the following non-uniqueness is known: considering solely the $%
Q $-divergence condition to first order (\ref{2.29}), it obviously remains
the freedom to add divergence and coboundary couplings to $W$ 
\begin{equation}
W^{(\mathbf{\beta},\mathbf{\gamma})}=W+\sum_l\beta _l\partial _{\nu }
D_l^{\nu }+\sum_j\gamma _j \{Q,K_j\},\quad \quad \beta _l,\gamma _j\in 
\mathbf{R},  \label{2.40}
\end{equation}
where $D_l^{\nu },K_j$ are restricted by (a)-(e). Taking additionally the $Q$%
-divergence condition to orders $n\geq 2$ (\ref{2.30}) into account, it
seems that this freedom can be maintained, due to the following result. It
is shown in \cite{D4} that the Q-divergence condition (\ref{2.30}) can be
satisfied to all orders (by choosing suitable normalizations) for any $(%
\mathbf{\beta},\mathbf{\gamma})$, if the 'generalized (free perturbative
operator) gauge invariance' \footnote{%
The 'generalized (free perturbative operator) gauge invariance' is the
following statement. To the interaction density $W\equiv W_{0}$ there exist
Wick polynomials $W_{1}^{\nu }$ and $W_{2}^{\mu \nu }$ with 
\begin{equation}
\lbrack Q,W_{0}]=i\partial _{\nu }W_{1}^{\nu },\quad \{Q,W_{1}^{\nu
}\}=i\partial _{\mu }W_{2}^{\mu \nu },\quad \lbrack Q,W_{2}^{\mu \nu }]=0 
\nonumber
\end{equation}
and the time ordered products of $W_{0},\,W_{1}^{\nu }$ and $W_{2}^{\mu \nu
} $ fulfil 
\begin{equation}
\lbrack Q,T(W_{j_{1}}(x_{1})...W_{j_{n}}(x_{n}))]_{\mp
}=i\sum_{l=1}^{n}\partial
^{x_{l}}T(W_{j_{1}}(x_{1})...W_{j_{l}+1}(x_{l})...W_{j_{n}}(x_{n})),\quad
j_{1},...,j_{n}\in \{0,1,2\},  \nonumber
\end{equation}
where we have the anticommutator on the l.h.s. iff $(j_{1}+...+j_{n})$ is
odd, and where $T(...W_{3}(x_{l})...)\equiv 0$ by definition. Similarly to (%
\ref{2.30}), the second equation is a normalization condition on the time
ordered products. A proof that it can be satisfied to all orders is still
missing in any nonabelian model. Nevertheless we strongly presume that this
statement holds true for all models which are BRST-invariant, especially for
the model studied below of three massive, selfinteracting vector fields.}
holds true for $(\mathbf{\beta},\mathbf{\gamma})=(\mathbf{0},\mathbf{0})$.
Moreover, under this assumption, one can prove 
\begin{equation}
P_{0}T(W^{(\mathbf{\beta},\mathbf{\gamma})}(x_{1})...W^{(\mathbf{\beta}, 
\mathbf{\gamma})}(x_{n}))P_{0}=P_{0}T(W^{(\mathbf{0},\mathbf{0})}(x_{1})
...W^{(\mathbf{0},\mathbf{0})}(x_{n}))P_{0}+\mathrm{divergences}
\label{2.43}
\end{equation}
(see \cite{D4}, 'divergences' means 'divergences of local operators') and
hence the physical $S$-matrix $S_{00}$ is independent from $(\mathbf{\beta},%
\mathbf{\gamma})$ (because the divergences vanish in the adiabatic limit in
pure massive theories).

Now we make the most general Ansatz for $W$ (up to divergence and coboundary
terms ) which satisfies (a)-(e) 
\begin{eqnarray}
W &=&f_{abc}:A_{a\,\mu }A_{b\,\nu }\partial ^{\nu }A_{c}^{\mu
}:+f_{abc}^{1}:u_{a}\partial ^{\mu }\tilde{u}_{b}A_{c\,\mu }:  \nonumber \\
&+&d_{abc}(:A_{a}^{\mu }\phi _{b}\partial _{\mu }\phi _{c}:-:A_{a}^{\mu
}\partial _{\mu }\phi _{b}\phi _{c}:)+e_{abc}:A_{a}^{\mu }A_{b\,\mu }\phi
_{c}:  \nonumber \\
&+&h_{abc}:\tilde{u}_{a}u_{b}\phi _{c}:+j_{abc}:\phi _{a}\phi _{b}\phi
_{c}:+k_{ab}(:HA_{a}^{\mu }\partial _{\mu }\phi _{b}:-:\partial _{\mu
}HA_{a}^{\mu }\phi _{b}:)  \nonumber \\
&+&l_{ab}:A_{a}^{\mu }A_{b\,\mu }H:+p_{ab}:H\tilde{u}_{a}u_{b}:+q_{ab}:H\phi
_{a}\phi _{b}:+r_{a}:H^{2}\phi _{a}:+s:H^{3}:  \nonumber \\
&+&(\mathrm{quadrilinear\>terms,\>e.g.}\>\sim :AAAA:,\,:AAu\tilde{u}%
:,\,:AA\phi \phi :,\,:H^{4}:)  \label{ansatzW}
\end{eqnarray}
where $f_{abc},\,f_{abc}^{1},\,d_{abc},\,e_{abc},\,h_{abc},\,j_{abc},%
\,k_{ab},\,l_{ab},\,p_{ab},\,q_{ab},\,r_{a},\,s\in \mathbf{R}$ are arbitrary
constants (which are introduced without any knowledge about a gauge group).
One can show that physical consistency (\ref{2.31}) requires that $f_{abc}$
is totally antisymmetric (i.e. this part of the result (\ref{2.62}) below is
independent of the restriction to \textit{three} vector fields which is made
in the following). Hence the simplest non-trivial model of selfinteracting
massive ($m_{a}>0\> \forall a$) vector fields is that of \textit{three}
fields: $a=1,2,3$ \footnote{%
The resulting model is usually called '$SU(2)$ Higgs-Kibble model'. It is
obtained from the electroweak theory, which is studied in detail in \cite{As}%
, by setting the Weinberg angle $\Theta _{W}=0$ and omitting the photon
field, which decouples in this case.}. From now on we specialize to this
case. In particular a single massive spin 1 field (corresponding to the
U(1)-case) cannot be selfinteracting.

Our aim is to determine the parameters in $W$ (\ref{ansatzW}) and the tree
normalization terms (see (\ref{2.49}) below) by the consistency condition (%
\ref{2.31}). The hope is that one can conclude from 
\begin{equation}
0=[Q,S_{n}]|_{\mathrm{ker}\,Q}=\frac{i^{n}}{n!}\int
dx_{1}...dx_{n}\,[Q,T(W(x_{1})...W(x_{n}))]|_{\mathrm{ker}\,Q}  \label{2.31n}
\end{equation}
that $[Q,T(W(x_{1})...W(x_{n}))]$ must be a sum of divergences of local
operators and that then one can follow the calculations in \cite{As}.

But there is a difficulty at first order, which is explained in detail in
appendix A. The adiabatic limit of the trilinear terms in $W$ vanishes
(except possibly the $H$-couplings) due to energy-momentum conservation.
This is just the fact that for stable particles there are no S-matrix
elements with three particle legs; the lowest tree contributions involve $%
\geq 4$ legs. Hence, we get no information about these terms from $%
[Q,S_{1}]|_{\mathrm{ker}\,Q}=0$. For the quadrilinear terms $W^{(4)}$ in $W$
the procedure works (see also appendix A): we obtain that $[Q,W^{(4)}]$ must
be the divergence of a Wick polynomial. In \cite{G} it is shown that this
implies $W^{(4)}=0$ except for a term $\sim :H^{4}:$. But the latter term
will be excluded by consistency (\ref{2.31}) to second order. (The
well-known $:H^{4}:$-coupling is of second order in $g$, it appears in the
framework of causal perturbation theory as a tree normalization term of $%
T(W(x_{1})W(x_{2}))$, see below.)

To determine the parameters in the trilinear terms $W^{(3)}$ of $W$ (\ref
{ansatzW}) we compute the tree diagrams in $T(W^{(3)}(x_{1})W^{(3)}(x_{2}))$
(which depend on these parameters) and require 
\begin{equation}
\int d^{4}x_{1}\,d^{4}x_{2}\,[Q,T(W^{(3)}(x_{1})W^{(3)}(x_{2}))|_{\mathrm{%
tree}}]|_{\mathrm{ker}\,Q}=0.  \label{2.60}
\end{equation}
Instead of the long explicit calculation we give a \textit{heuristic}
description:

$\bullet $ For the terms with $x_{1}\not=x_{2}$ the $T$-product factorizes,
e.g. for $x_{1}\not\in x_{2}+\bar{V}^{-}$ we have 
\begin{equation}
\lbrack Q,T(W^{(3)}(x_{1})W^{(3)}(x_{2}))|_{\mathrm{tree}%
}]=[Q,W^{(3)}(x_{1})]W^{(3)}(x_{2})|_{\mathrm{tree}%
}+W^{(3)}(x_{1})[Q,W^{(3)}(x_{2}))]|_{\mathrm{tree}}.  \label{2.61}
\end{equation}
This makes it plausible that the cancelation of these terms\footnote{%
The fact that the terms $x_{1}\not=x_{2}$ cannot be canceled by diagonal
terms $x_1=x_2$ becomes clear from the explicit expressions.} yields the
same restrictions as the $Q$-divergence condition to first order (\ref{2.29}%
): $[Q,W^{(3)}]=i\partial _{\nu }W_{1}^{(3)\,\nu }$ for some Wick monomial $%
W_{1}^{(3)\,\nu }$.\footnote{%
It is obvious that the $Q$-divergence condition to first order implies the
cancelation of the terms $x_{1}\not=x_{2}$ in (\ref{2.60}), but here we
proceed in the opposite direction.} In this way we obtain 
\begin{eqnarray}
f_{abc} &=&-f_{abc}^{1}\quad \mathrm{and}\quad f_{abc}\quad \mathrm{%
is\>totally\>antisymmetric},  \nonumber \\
d_{abc} &=&f_{abc}\frac{m_{b}^{2}+m_{c}^{2}-m_{a}^{2}}{4m_{b}m_{c}},\quad
e_{abc}=f_{abc}\frac{m_{b}^{2}-m_{a}^{2}}{2m_{c}},  \nonumber \\
h_{abc} &=&f_{abc}\frac{m_{a}^{2}+m_{c}^{2}-m_{b}^{2}}{2m_{c}},\quad
j_{abc}=0,\quad r_{a}=0  \label{2.62}
\end{eqnarray}
and some relations for $k_{ab},\,l_{ab},\,p_{ab}$ and $q_{ab}$ (\cite{As}, 
\cite{G} and \cite{Sch}). There results no restriction on $s$. One easily
verifies that these values of the parameters are not only necessary for the
cancelation of the terms $x_1\not= x_2$ in (\ref{2.60}) and for (\ref{2.29}%
), they are also sufficient. By absorbing a constant factor in $g$ we obtain 
\begin{equation}
f_{abc}=\epsilon _{abc}.  \label{f=e}
\end{equation}
The latter are the structure constants of $su(2)$. So the gauge group
structure is not put in, it comes out as a consequence of physical
consistency and the Ansatz (\ref{ansatzW}) for $W$. For more complicated
models (i.e. more than three vector fields) this conclusion is impossible at
this stage, because one does not know that the $f_{abc}$'s satisfy the
Jacobi identity. But then the latter is obtained in the next step (see
below). So far it is possible that all couplings involving the $H$-field
vanish, i.e. the Higgs field is not yet needed.

$\bullet$ The remaining terms in (\ref{2.60}) come from the diagonal $%
x_1=x_2 $. Their cancelation cannot be achieved without $H$-couplings. This
condition yields important results:\footnote{%
These results agree precisely with the ones derived from the $Q$-divergence
condition (for second order tree diagrams) in \cite{As}, \cite{Sch} and with 
\cite{G}.}

- The $f_{abc}$'s must fulfil the Jacobi identity. (In our simple model this
is already known (\ref{f=e}).) \footnote{%
Stora \cite{St} found (for an arbitrary number of massless selfinteracting
vector fields) that the $Q$-divergence condition to first order implies that
the coupling parameters are totally antisymmetric and that the $Q$%
-divergence condition for second order tree diagrams yields the Jacobi
identity.}

- The masses must agree 
\begin{equation}
m\equiv m_{1}=m_{2}=m_{3}.  \label{2.47}
\end{equation}

- The $H$-coupling parameters take the values 
\begin{equation}
k_{ab}=\frac{\kappa }{2}\delta _{ab},\quad l_{ab}=-\frac{\kappa m}{2}\delta
_{ab},\quad p_{ab}=\frac{\kappa m}{2}\delta _{ab},\quad q_{ab}=\frac{\kappa
m_{H}^{2}}{4m}\delta _{ab}  \label{2.48}
\end{equation}
where $\kappa \in \{-1,1\}$. The parameter $s$ is still free.

- There is no term $\sim :H^4:$ in $W$ (i.e. in first order in $g$).

- In $T(W(x_{1})W(x_{2}))\vert_{\mathrm{tree}}$ the C-number distributions
are Feynman propagators with derivatives: $\Delta
^{F}(x_{1}-x_{2}),\>\partial _{\mu }\Delta ^{F}(x_{1}-x_{2})$ and $\partial
_{\nu }\partial _{\mu }\Delta ^{F}(x_{1}-x_{2})$. The first two extend
uniquely to the diagonal $x_{1}=x_{2}$ and the last one has a distinguished
extension, namely $\partial _{\nu }\partial _{\mu }\Delta ^{F}(x_{1}-x_{2})$ 
\footnote{%
The general extension which is Poincar\'{e} covariant (D) and does not
increase the scaling degree (F) reads: $\partial _{\nu }\partial _{\mu
}\Delta ^{F}(x_{1}-x_{2})+Cg_{\nu \mu }\delta (x_{1}-x_{2}),\quad C\in 
\mathbf{C}$ arbitrary.}. We denote this extension by $T(W(x_{1})W(x_{2}))|_{%
\mathrm{tree}}^{0}$. So-called 'tree normalization terms' 
\begin{equation}
N(x_{1},x_{2})=C\delta
(x_{1}-x_{2}):B_{1}(x_{1})B_{2}(x_{2})B_{3}(x_{3})B_{4}(x_{4}):,\quad \quad
C\in \mathbf{R}\>\mathrm{or}\>i\mathbf{R}  \label{2.49}
\end{equation}
($B_{1},...,B_{4}\in \{A^\mu,\,u,\,\tilde u,\,\phi,\,H\}$) can be added to $%
T(W(x_{1})W(x_{2}))|_{\mathrm{tree}}^{0}$, if they satisfy the properties
(b) (Poincar\'{e} covariance), (c) (ghost number), (d) (scaling degree) and
(e) (unitarity) which are required above for $W$ (here they restrict $%
N(x_{1},x_{2})$). These tree normalization terms (\ref{2.49}) correspond to
the quadrilinear terms of order $g^{2}$ ($g$ denotes the coupling constant)
in the interaction Lagrangian of the conventional theory, they have the same
influence on the perturbation series of the $S$-matrix. \textit{The
cancelation of the terms $x_1=x_2$ in (\ref{2.60}) fixes the possible tree
normalization terms} (i.e. the constants $C$ in (\ref{2.49})) \textit{%
uniquely} (in terms of $s$), except for \footnote{%
For the expert we mention that tree normalization terms with $B_1,...,B_4$
exclusively scalar fields (e.g. $N_{H^4}$) are required. In contrast to the
other tree normalization terms (i.e. the tree normalization terms with
vector field factors) they violate the normalization condition \textbf{(N3)}
in \cite{D-F} (or (43) in \cite{E-G}). But this is no harm.} 
\begin{equation}
N_{H^4}(x_{1},x_{2})=\lambda \delta (x_{1}-x_{2}):H^{4}(x_{1}):,\quad \quad
\lambda \in \mathbf{R}.  \label{2.50}
\end{equation}

$\bullet$ The parameters $s$ and $\lambda $, which are still free, are
determined by physical consistency (\ref{2.31n}) for the tree diagrams to 
\textit{third} order (analogously to \cite{As}): 
\begin{equation}
s=\frac{m_{H}^{2}}{4m},\quad \quad \quad \lambda =-\frac{m_{H}^{2}}{16m^{2}}.
\label{2.51}
\end{equation}

So $W$ and the tree diagram normalizations to second order are completely
determined (up to the sign $\kappa $, which is conventional) and these terms
agree precisely with the interaction Lagrangian obtained by the Higgs
mechanism. We have made the assumption that there is at most one physical
scalar field $H$. This assumption is most probably not necessary. By
specializing the results of \cite{Sch} to our model (of three
selfinteracting massive vector fields) one finds that there is no solution
of the $Q$-divergence condition with more than one $H$-field, provided the $%
H $-fields couple to the vector or ghost fields. 

The Higgs potential is not put in here, it is derived from physical
consistency. Spontaneous symmetry breaking plays no role in this approach,
because we \textit{start with the massive free incoming fields}. A proof
that physical consistency can be satisfied to all orders (by choosing
suitable normalizations) is missing up to now, but we are convinced that
this holds true.

\section{Renormalizability and Ghosts}

The formulation of massive selfinteracting vector mesons of the previous
section will now serve as a point of departure for a more fundamental
conceptual discussion. Similar to Weinberg (Weinberg's old work on Feynman
rules for higher spin) we start with Wigner's theory of particle
representations; in our case because we want to avoid any parallelism to
(quasi)classical systems (quantization) and (for reasons which will become
gradually clear to the reader) develop local quantum physics from an
intrinsic point of view as much as possible. In particular we would like to
understand the curious phenomenon that, contrary to the classical situation,
the possibilities of perturbative renormalizable QFT are the more
restrictive, the higher the spin. Whereas the classical theory is in need of
an additional selection principle (the gauge principle in case of zero
mass), local quantum physics for spin$\geq 1$ is more restrictive: \textit{%
the particle content and renormalizability fix the vector meson theory}
(where the possible triviality for spin $>1$ is a special case\footnote{%
In a recent paper Scharf and Wellmann \cite{SW} have shown that there exists
no renormalizable theory for s=2 which satisfies the free perturbative
operator gauge invariance.}). It is this result red backward into
(quasi)classical field theory, which, in the spirit of Bohr's correspondence
principle gives a fundamental physical support for the \textit{classical
gauge selection principle} (which then gives the strong link with the
mathematical-aesthetical appeal to fibre bundles). As a side result we will
learn that the massive theory fulfills the Schwinger-Swieca \cite{Schw} \cite
{Sw} screening mechanism, and that (similar to the gauge interpretation in
terms of a Higgs field) the theory has more physical degrees of freedom than
the massive vector mesons from which we started in zeroth order, namely
consistent perturbation theory requires the introduction of a scalar $H$%
-field (the Higgs field without Higgs condensate\footnote{%
We use the notation $H$-field only in order to avoid any association with
Higgs condensates.}) of the previous section.

It is well-known that the step from the Wigner representation theory of
particles (positive energy irreducible representations of the Poincar\'{e}
group with finite spin/helicity) to local free fields is described in terms
of intertwiners $u,v$\footnote{%
The letter $u$, which was used in the previous section for a ghost field,
means here an intertwiner. The ghost fields in the Wigner one-particle space
will be denoted by $(\omega ,\bar{\omega},\varphi )$. The corresponding Fock
space fields are the fields $(u,\tilde{u},\phi )$ of the previous section.} 
\begin{equation}
\psi ^{\left[ n_{+},n_{-}\right] }(x)=\int \sum_{s_{3}=-s}^{s}\left\{
e^{-ipx}u(\vec{p},s_{3})a(\vec{p},s_{3})+e^{ipx}v(\vec{p},s_{3})b^{\ast }(%
\vec{p},s_{3})\right\} \frac{d^{3}p}{2\omega }  \label{fields}
\end{equation}
which intertwine the Wigner representation matrices $D^{(s)}(R(\Lambda ,p))$
with matrices of the covariant Lorentz group representation $%
D^{[n_{+},n_{-}]}(\Lambda )$ 
\begin{equation}
D^{\left[ n_{+},n_{-}\right] }(\Lambda )u(p)=u(\Lambda p)D^{(s)}(R(\Lambda
,\Lambda p))
\end{equation}
where for convenience we have collected the $(2s+1)$ $u^{(n_{+},n_{-})}(p)$
mixed (un)dotted $(2n_{+}+1)(2n_{-}+1)$ component u-spinors into a
rectangular $(2n_{+}+1)(2n_{-}+1)\times (2s+1)$ matrix $u(p)$ and similar
for the $v^{\prime }s.$

Let us first note that the covariantized inner product in the \textit{%
one-particle Wigner space} for $s\geq 1$ contains necessarily first or
higher powers of momenta. Associated with this is the fact that the
intertwiners $u,v$ have a dimension $\geq 1$ which immediately translates
into an operator dimension of the field dim$\psi \geq 2.$ Since interaction
densities $W_{0}\equiv W$ are at least trilinear in free fields and since
the smallest possible operator dimension is 1 (for scalar fields), it is
impossible to satisfy the renormalizability condition dim$W\leq 4$ within
the St\"{u}ckelberg-Bogoliubov-Epstein-Glaser operator approach. Inspired by
the idea of a cohomological representation of the physical space and the
physical observables in the previous section one looks for a cohomological
extension of the Wigner space for massive vector mesons in order that the
associated two-point function (or propagation kernel) has a milder
(renormalizable) high momentum behavior. The results of the previous section
also suggest the simplest possibility to achieve that, namely to use three
additional indefinite metric scalar wave functions (where on the Wigner
level the ``statistics'' is yet undetermined). More precisely we form an
extended Hilbert space $H_{ext}$ by the multicomponent wave functions $%
(A_{\mu },\omega ,\bar{\omega},\varphi )$ defined on the mass shell. $%
H_{ext} $ has in addition to a positive definite inner product another one
which does not have this property and corresponds to the Krein structure of
the previous section. The latter is relevant in order to have at least some
pseudo-unitary Lorentz covariance and a definition of modular localized
subspaces. On $H_{ext}$ we define a BRS-like operator by 
\begin{equation}
s_{W}\left( 
\begin{array}{ll}
A_{\mu }^{a}(p) &  \\ 
\omega _{a}(p) &  \\ 
\bar{\omega}_{a}(p) &  \\ 
\varphi _{a}(p) & 
\end{array}
\right) =\left( 
\begin{array}{ll}
p_{\mu }\omega _{a}(p) &  \\ 
0 &  \\ 
-p^{\mu }A_{\mu }^{a}(p)-im_{a}\varphi _{a}(p) &  \\ 
im_{a}\omega _{a}(p) & 
\end{array}
\right) .  \label{sW}
\end{equation}
The so defined $s_{W}$-operation defines a differential space since the
definition easily leads to $s_{W}^{2}=0.$ One then uses this $s_{W}$ in
order to write the following cohomological representation for the physical
(Wigner) Hilbert space $H_{W}$ in terms of the above extended space $H_{ext}$
\begin{eqnarray}
H_{W} &=&\frac{\mathrm{ker}\,s_{W}}{\mathrm{ran}\,s_{W}} \\
&=&cl.\left\{ A_{\mu }(p)\mid p^{\mu }A_{\mu }(p)=0,-\int A_{\mu }(p)A^{\mu
}(p)\frac{d^{3}p}{2\omega }<\infty \right\}  \nonumber \\
&=&cl.\left\{ A_{\mu }(p)\mid -\int A_{\mu }(p)(g^{\mu \nu }-\frac{p^{\mu
}p^{\nu }}{m^{2}})A_{\nu }(p)\frac{d^{3}p}{2\omega }<\infty \right\} 
\nonumber
\end{eqnarray}
i.e. we obtain the $L^{2}$-closure of the space of transversal vector wave
functions which in terms of the associated fields (\ref{fields}) had the
high dimension dim$A=2$ which was responsible for the lack of renormalizable
interactions within the original (non-extended) formulation. (The
transversality condition does not lower the dimension of the vector wave
functions or the corresponding fields.) On the other hand the extended
Hilbert space has no transversality condition and obeys the classical
assignment of dimensions (i.e. dim$A=1$ in $H_{ext}$). This is due to the
fact that ghost contributions damp the high momentum behavior of the
associated two-point function.\footnote{%
In the Lagrangean framework there is an alternative method to lower the
dimension of the massive vector field $A$ from $2$ to $1$. In contrast to
massless vector fields, the Proca field ($\mathcal{L}= -\frac{1}{4}F^{2}+%
\frac{1}{2} m^{2}A^{2},\> F^{\mu \nu }\equiv \partial ^{\mu }A^{\nu
}-\partial ^{\nu}A^{\mu }$) has a well-defined propagator without
introducing a ``gauge fixing term'', but it has dimension dim$A=2$.
St\"{u}ckelbergs trick is to add a ``gauge fixing term'' $\mathcal{L}_{1}=-%
\frac{1}{2}\lambda (\partial \cdot A)^{2},\>\lambda >0$ (this misleading
terminology is chosen because because $\mathcal{L}_{1}$ has the same form as
the gauge fixing Lagrangean in the massless case), which damps the high
momentum behavior of the propagator such that dim$A=1$. In this
St\"{u}ckelberg formalism, the necessity to introduce ghosts shows up in the
fact that without ghosts it is impossible to define a \textit{stable}
physical subspace or factor space. For $\lambda =1$ (Feynman gauge) (\ref
{2.7}) one obtains the same free vector fields as in the quantization of the
cohomological extension of the Wigner one-particle space. We prefer the
latter method, because we do not want to rely on a Lagrangean framework,
instead we want to be close to the particle picture.}

The extension has however an influence on the modular localization theory in
Wigner space where the latter is (via the CCR/CAR functor) the preempted
locality in Fock space. For example the wedge (Rindler or
Bisognano-Wichmann) localized subspace involves instead of the simple
complex-conjugation in addition the Krein operator $\eta $ i.e. one has for
the action of the pre-Tomita operator\footnote{%
The pre-Tomita operator $S$ has nothing to do with the above nilpotent
operators $s_{W}$ (\ref{sW}) (in the one particle Wigner space) or $s$ (\ref
{s}) in Fock space.} 
\begin{equation}
(S\psi )(p)=(\eta \overline{\psi })(p).
\end{equation}
Where $\psi $ is the multi-component wave function involving the ghosts in
addition to the vector potential. The standard modular localization theory
can be found in \cite{Schr1} and the adaptation to the present extended
``pseudo-modular'' case will be treated in a separate paper. The physical
Wigner subspace (more precisely it is a cohomologically defined factor space)
is precisely characterized by the validity of the ``correct'' modular
localization associated with the Tomita theory. Although our ghost extension
of the (m,s=1) Wigner representation is not uniquely fixed (we chose a
``minimal'' extension) we believe that any other cohomological extension
which also lowers the dim$A=2$ down to its classical value dim$A=1$ will
contain the minimal and possible additional pieces which do not change the
physical content\footnote{%
The geometrical Faddeev-Popov method can also be considered as a minimal
extension of the functional measure. It owes its unique appearance more to
geometric than quantum physical reasoning.}.

The next step from particles to fields is the answer to the question of what
is the action of $s_{W}$ on the \textit{multi particle tensor space} \cite{DV}%
. From the usual Fock space formalism we are used to the following action of
derivations $\delta $ on tensor products 
\[
\delta (\psi _{1}\otimes \psi _{2})=\delta \psi _{1}\otimes \psi _{2}+\psi
_{1}\otimes \delta \psi _{2} 
\]
It is easy to see that the tensor product action of $s_{W}$ must include a
grading in order to maintain the nilpotency 
\begin{equation}
s(\psi _{1}\otimes \psi _{2})=s\psi _{1}\otimes \psi _{2}+\psi _{1}(-)^{%
\mathrm{degree}\>\psi _{1}}\otimes s\psi _{2},  \label{s}
\end{equation}
where $s$ denotes the Fock space version of $s_{W}$.\footnote{%
>From the context it should be clear whether we mean by the letter $s$ the
present nilpotent Fock space operator $s$ or the spin.} So, two of the three
scalar ghost fields $(u,\tilde{u},\phi )$\footnote{%
We recall that $(u,\tilde{u},\phi )$ are the Fock space fields of the
previous section and correspond to $(\omega ,\bar{\omega},\varphi )$ in the
Wigner one-particle space.}, which are companions of each massive vector
meson field, are required to be graded fermionic fields and the third one
must be bosonic.

In this way we obtain the Fock space formalism of the previous section with $%
\left[ Q,\cdot \right] $ being the implementation of $s$ in Fock space.
Whereas the ghost formalism can be pursued back into the Wigner one-particle
theory, the necessity to choose trilinear couplings at first order in $g$
with coefficients fulfilling group theoretical symmetry, as well as the
necessity of enlargement of the vector meson setup by additional physical
degrees of freedom (whose simplest and perhaps only realization are the
scalar $H$-fields) only shows up as a consistency requirement above the
zeroth order.

\section{Determination by Field Content}

In the second section physical consistency was formulated for the $S$-matrix
(\ref{2.31}). We are now looking for a corresponding condition in terms of 
\textit{interacting fields}. Such fields (including composites) are defined
by means of the Bogoliubov transition functional $S(\mathbf{g})$ (\ref{2.1})
in Fock space as formal power series in $g_{0}$. The interacting field $%
W_{j\,\mathrm{int}}(x;g_{0}W)$ due to the interaction $W\equiv W_{0}$ and
corresponding to the Wick polynomial $W_{j},\>j=1,...,G$ of free fields, is
defined by 
\begin{equation}
W_{j\,\mathrm{int}}(x;g_{0}W)\equiv \frac{\delta }{i\delta g_{j}(x)}%
S(g_{0},0,...,0)^{-1}S(g_{0},0,...,g_{j},0,...)|_{g_{j}=0}.  \label{int}
\end{equation}
By inserting (\ref{2.1}) one obtains the perturbative expansion of the
interacting fields 
\begin{equation}
W_{j\,\mathrm{int}}(x;g_{0}W)=W_{j}(x)+\sum_{n=1}^{\infty }\frac{i^{n}}{n!}%
\int
d^{4}x_{1}...d^{4}x_{n}%
\,g_{0}(x_{1})...g_{0}(x_{n})R(W(x_{1})...W(x_{n});W_{j}(x)),
\label{intentw}
\end{equation}
with the 'totally retarded products' 
\begin{equation}
R(A_{1}(x_{1})...A_{n}(x_{n});A(x))\equiv \sum_{I\subset
\{1,...,n\}}(-1)^{|I|}\bar{T}(A_{i}(x_{i}),\,i\in I)T(A_{k}(x_{k}),\,k\in
I^{c},A(x)),  \label{retprod}
\end{equation}
where $A_{1},...,A_{n},A$ are Wick polynomials, $I^{c}\equiv
\{1,...,n\}\setminus I$ and $\bar{T}$ denotes the 'anti-chronological
product'. The corresponding generating functional is $S(\mathbf{g})^{-1}$.
The anti-chronological products can be obtained uniquely from the time
ordered products by the usual inversion of a formal power series 
\begin{equation}
\bar{T}(A_{1}(x_{1})...A_{n}(x_{n}))=\sum_{P\in \mathrm{Part}%
\,\{1,...,n\}}(-1)^{|P|+n}\prod_{p\in P}T(A_{i}(x_{i}),\,i\in p).
\label{antiT}
\end{equation}
By means of causality (\ref{2.4}) one easily finds that the $R$-products (%
\ref{retprod}) have totally retarded support with respect to the
distinguished coordinate x. 
\begin{equation}
\mathrm{supp}\>R(A_{1}(x_{1})...A_{n}(x_{n});A(x))\subset
\{(x_{1},...x_{n};x)\mid x_{i}\in x+\bar{V}_{-},\,\forall i=1,...n\}.
\label{suppR}
\end{equation}
In a pure massive theory the strong adiabatic limit of the interacting
fields exists as a formal power series in $g:=g_{0}(0)$: 
\begin{equation}
W_{j\,\mathrm{int}}(x)\psi \equiv W_{j\,\mathrm{int}}(x;W)\psi \equiv
\lim_{\epsilon \rightarrow 0}W_{j\,\mathrm{int}}(x;g_{0\,\epsilon }W)\psi
,\quad \mathrm{where}\quad g_{0\,\epsilon }(x)\equiv g_{0}(\epsilon
x),\>\psi \in \mathcal{D},  \label{adlim}
\end{equation}
because this holds true for $S(\mathbf{g})$ (\ref{2.6}) \cite{E-G1}. \vskip%
0.3cm A $\ast $-symmetrical interacting field in the adiabatic limit (i.e. $%
\phi _{\mathrm{int}}(x)\equiv \phi _{\mathrm{int}}(x;W)=\lim_{\epsilon
\rightarrow 0}\phi _{\mathrm{int}}(x;g_{0\,\epsilon }W),\>\phi _{\mathrm{int}%
}^{\ast }=\phi _{\mathrm{int}}$) is called \textit{physical} (or an \textit{%
observable}) if it induces a well-defined operator on the factor space $%
\mathcal{H}_{\mathrm{phys}}^{\prime }$ (\ref{2.22}). This holds true iff 
\begin{equation}
\phi _{\mathrm{int}}(f;W)\>\mathrm{ker}\,Q\subset \mathrm{ker}\,Q\quad
\wedge \quad \phi _{\mathrm{int}}(f;W)\>\mathrm{ran}\,Q\subset \mathrm{ran}%
\,Q,\quad \forall f\in \mathcal{S}(\mathbf{R}^{4})  \label{physical}
\end{equation}
(cf.(\ref{2.25})). This is equivalent to 
\begin{equation}
\lbrack Q,\phi _{\mathrm{int}}(f;W)]|_{\mathrm{ker}\,Q}=0,\quad \quad
\forall f\in \mathcal{S}(\mathbf{R}^{4}).  \label{physical1}
\end{equation}
(The nontrivial part of this statement is that (\ref{physical1}) implies $%
\phi _{\mathrm{int}}(f)\>\mathrm{ran}\,Q\subset \mathrm{ran}\,Q$. However,
using the notations of sect. 2, the condition (\ref{physical1}) is
equivalent to $\phi _{\mathrm{int}}(f)_{+-}=0=\phi _{\mathrm{int}}(f)_{+0}$.
Hence $\phi _{\mathrm{int}}(\bar{f})_{0-}=(\phi _{\mathrm{int}%
}(f)_{+0})^{\ast }=0$. Together we obtain $\phi _{\mathrm{int}}(f)\>\mathrm{%
ran}\,Q\subset \mathrm{ran}\,Q$.)

Let $F^{\mu\nu}_a\equiv\partial^\mu A^\nu_a-\partial^\nu A^\mu_a$. We now 
\textit{require that (for each $a=1,...,M$) there exists a physical field} $%
\mathcal{F}^{\mu\nu}_{a\,\mathrm{int}}$, i.e. 
\begin{equation}
[Q,\mathcal{F}^{\mu\nu}_{a\,\mathrm{int}}(x;W)]\vert_{\mathrm{ker}\,Q}=0,
\label{QF}
\end{equation}
with the additional properties:

(i) $\mathcal{F}^{\mu\nu}_{a\,\mathrm{int}}=F^{\mu\nu}_{a\,\mathrm{int}}+
\sum_kc_k\psi^{\mu\nu}_{ak\,\mathrm{int}}$, where the $c_k$ are formal power
series of (constant) complex numbers and $\mathcal{F}^{\mu\nu}_{a\,\mathrm{%
int}}$ agrees in zeroth order with the free $F^{\mu\nu}_a$, i.e. the $c_k$
vanish to zeroth order: $c_k^{(0)}=0$;

(ii) the zeroth order $\psi^{\mu\nu}_{ak}$ of $\psi^{\mu\nu}_{ak\, \mathrm{%
int}}$ is a Wick monomial which has precisely one factor $F$ or $A$, the
other factors are ghost or scalar fields, and $\psi^{\mu\nu}_{ak}\not=F^{\mu%
\nu}_b,\>\forall a,b\in\{1,...,M\}$;

(iii) the scaling degree (or mass dimension) of $\psi_{ak}$ is $\leq 4$;

(iv) $\mathcal{F}^{\mu\nu}_{a\,\mathrm{int}}$ is a Lorentz tensor of second
rank and is anti-symmetrical in $(\mu,\nu)$;

(v) the ghost number of $\mathcal{F}^{\mu\nu}_{a\,\mathrm{int}}$ is zero;

(vi) $\mathcal{F}^{\mu\nu}_{a\,\mathrm{int}}(x;W)^*= \mathcal{F}%
^{\mu\nu}_{a\,\mathrm{int}}(x;W)$.

The requirement (ii) is badly motivated (except the demand that $\psi_{ak}$
must be a Wick monomial), its main purpose is to shorten the calculations.
However, the condition $\psi^{\mu\nu}_{k}\not=F^{\mu\nu}$ in (ii) is
necessary for the uniqueness of $\mathcal{F}_{\mathrm{int}}$: the fields 
\begin{equation}
(1+\sum_{k=1}^\infty b_k g^k)\mathcal{F}^{\mu\nu}_{a\,\mathrm{int}},\quad
\quad b_k=\mathrm{const.}\in \mathbf{R}  \nonumber
\end{equation}
satisfy all other requirements if $\mathcal{F}_{\mathrm{int}}$ does so. It
is an interesting question how far (ii) can be weakened such that the
uniqueness of $\mathcal{F}_{\mathrm{int}}$ does not get lost.

Due to the normalization condition (F) (scaling degree) for the time ordered
products, the property (iii) implies dim$\mathcal{F}^{\mu\nu}_{a\,\mathrm{int%
}}\leq 4$. To specify (iv), (v) and (vi) note that they must be fulfilled in
particular by the Wick monomials $\psi_{ak}$ (in the case of (vi) the
coefficients $c_k$ are also involved: $c_k^*\psi_{ak}^*=c_k\psi_{ak}$).
However, these three requirements also restrict the normalization of the
higher orders of $\psi^{\mu\nu}_{ak\,\mathrm{int}}$ and $F^{\mu\nu}_{a\,%
\mathrm{int}}$. It is easy to see that these additional normalization
conditions can be fulfilled (e.g. by antisymmetrization in $(\mu,\nu)$ of an
arbitrary Poincar\'{e} covariant extension) and we assume that the
normalizations are always done in such a way.

The most general Ansatz which is compatible with (i)-(vi) reads 
\begin{eqnarray}
\mathcal{F}_{d\,\mathrm{int}}^{\mu \nu } &=&F_{d\,\mathrm{int}}^{\mu \nu
}+t_{dab}(F_{a}^{\mu \nu }\phi _{b})_{\mathrm{int}}+\tilde{t}%
_{dab}(A_{a}^{\mu }\partial ^{\nu }\phi _{b}-A_{a}^{\nu }\partial ^{\mu
}\phi _{b})_{\mathrm{int}}  \nonumber \\
+v_{dabc}( &:&F_{a}^{\mu \nu }\phi _{b}\phi _{c}:)_{\mathrm{int}}+\tilde{v}%
_{dabc}(:A_{a}^{\mu }\partial ^{\nu }\phi _{b}\phi _{c}:-:A_{a}^{\nu
}\partial ^{\mu }\phi _{b}\phi _{c}:)_{\mathrm{int}}  \nonumber \\
+w_{dabc}( &:&F_{a}^{\mu \nu }u_{b}\tilde{u}_{c}:)_{\mathrm{int}}+\tilde{w}%
_{dabc}(:A_{a}^{\mu }u_{b}\partial ^{\nu }\tilde{u}_{c}:-:A_{a}^{\nu
}u_{b}\partial ^{\mu }\tilde{u}_{c}:)_{\mathrm{int}}  \nonumber \\
+w_{dabc}^{\prime }( &:&A_{a}^{\mu }\partial ^{\nu }u_{b}\tilde{u}%
_{c}:-:A_{a}^{\nu }\partial ^{\mu }u_{b}\tilde{u}_{c}:)_{\mathrm{int}} 
\nonumber \\
&&+x_{da}(F_{a}^{\mu \nu }H)_{\mathrm{int}}+\tilde{x}_{da}(A_{a}^{\mu
}\partial ^{\nu }H-A_{a}^{\nu }\partial ^{\mu }H)_{\mathrm{int}}  \nonumber
\\
&&+y_{dab}(F_{a}^{\mu \nu }\phi _{b}H)_{\mathrm{int}}+\tilde{y}%
_{dab}(A_{a}^{\mu }\partial ^{\nu }\phi _{b}H-A_{a}^{\nu }\partial ^{\mu
}\phi _{b}H)_{\mathrm{int}}  \nonumber \\
&&+y_{dab}^{\prime }(A_{a}^{\mu }\phi _{b}\partial ^{\nu }H-A_{a}^{\nu }\phi
_{b}\partial ^{\mu }H)_{\mathrm{int}}  \nonumber \\
+z_{da}( &:&F_{a}^{\mu \nu }H^{2}:)_{\mathrm{int}}+\tilde{z}%
_{da}(:A_{a}^{\mu }\partial ^{\nu }HH:-:A_{a}^{\nu }\partial ^{\mu }HH:)_{%
\mathrm{int}},  \label{ansatzF}
\end{eqnarray}
where $t_{dab},...,w_{dabc}^{\prime },x_{da},...,\tilde{z}_{da}$ are
(arbitrary) constants which are formal power series in $\mathbf{R}$ 
\begin{equation}
t_{dab}=\sum_{k=1}^{\infty }t_{dab}^{(k)}g^{k},\quad \quad
x_{da}=\sum_{k=1}^{\infty }x_{da}^{(k)}g^{k},\quad \quad \mathrm{etc.}. 
\nonumber
\end{equation}
We define 
\begin{equation}
\mathcal{F}_{d}^{\mu \nu \,(k)}\equiv t_{dab}^{(k)}F_{a}^{\mu \nu }\phi
_{b}+...+x_{da}^{(k)}F_{a}^{\mu \nu }H+...\quad \quad \forall k\geq 1, 
\nonumber
\end{equation}
hence\footnote{%
By definition the map $\phi \rightarrow \phi _{\mathrm{int}}$ ($\phi $ a
Wick monomial) is linear with $C^{\infty }$-functions as coefficients (here
the coefficients are constants).} 
\begin{equation}
\mathcal{F}_{d\,\mathrm{int}}^{\mu \nu }=F_{d\,\mathrm{int}}^{\mu \nu
}+\sum_{k=1}^{\infty }\mathcal{F}_{d\,\mathrm{int}}^{\mu \nu \,(k)}g^{k}.
\label{ReiheF}
\end{equation}
By definition we may assume $v_{dabc}=v_{dacb}$.

We require that the interaction density $W$ satisfies the properties (a)-(e)
listed in section 2. The physical consistency of the $S$-matrix $[Q,S]
\vert_{\mathrm{ker}\,Q}=0$ is omitted here, we replace it by $[Q, \mathcal{F}%
_{\mathrm{int}}(x;W)]=0$ (see (\ref{QF'}) below). In particular we require
dim$W\leq 4$ which ensures renormalizability. Additionally we assume that $W$
contains no quadrilinear terms. Most probably this assumption is not
necessary, i.e. the other requirements (including the physicality (\ref{QF}%
)) exclude the quadrilinear terms. But this still needs to be checked and
will be left open here.

We now replace the requirement (\ref{QF}) by 
\begin{equation}
\lbrack Q,\mathcal{F}_{a\,\mathrm{int}}^{\mu \nu }(x;W)]=0.  \label{QF'}
\end{equation}
This may be justified in the following way\footnote{%
We thank Klaus Fredenhagen for bringing this argument to our attention.}.
Since we are working in the adiabatic limit our $Q$ (\ref{2.36})
(constructed in terms of incoming free fields) agrees with the Kugo-Ojima
operator $Q_{\mathrm{int}}$ \cite{K-O}, \cite{D-F} which implements the
BRST-transformation of the interacting fields. Hence $[Q,\mathcal{F}_{a\,%
\mathrm{int}}^{\mu \nu }(x;W)]$ is again a local operator. But by the
Reeh-Schlieder theorem \cite{R-S} such an operator is zero if it vanishes on
the vacuum, which is an element of the kernel of $Q$. Thus the requirements (%
\ref{QF}) and (\ref{QF'}) are equivalent.\footnote{%
This argument cannot be applied to $[Q,S]|_{\mathrm{ker}\,Q}=0$ (\ref{2.31}%
), because the $S$-matrix is non-local.} \vskip0.3cm The main (completely
new) result of this section is that the \textit{physicality (\ref{QF'})
fixes the parameters $t_{dab},...,\tilde{z}_{da}$ in the Ansatz (\ref
{ansatzF}) and the interaction $W$} (including the tree normalization terms
to second order (\ref{2.49})) up to the same non-uniqueness as in section 2,
namely the addition of divergence- and coboundary couplings to $W$ (\ref
{2.40}). Especially we will see that \textit{an additional physical degree
of freedom is required: the Higgs field $H$ is needed in the interaction $W$
as well as in the field} $\mathcal{F}_{d\,\mathrm{int}}^{\mu \nu }$. \vskip%
0.3cm We verify this statement in appendix B by explicit calculation of the
tree diagrams to lowest orders. We do this only for the simplest non-trivial
model, namely three selfinteracting massive vector fields ($m_{a}>0,\>a=1,
2,3$) as in section 2. We make the same Ansatz for $W$ as in section 2 , but
without the quadrilinear terms. Up to divergence- and coboundary couplings (%
\ref{2.40}) this is the most general trilinear Ansatz which fulfills the
requirements (a)-(e) of section 2.

The parameters in the Ansatz for $W$ and $\mathcal{F}_{d\,\mathrm{int}}^{\mu
\nu }$ (\ref{ansatzF}) are determined by inserting these expressions into
the physicality requirement (\ref{QF'}). Here we only state the results:

- for $W$ we obtain precisely the same expression as in section 2,

- we have computed the parameters in $\mathcal{F}_{\mathrm{int}}$ (which are
formal power series) up to second order in $g$ 
\begin{eqnarray}
\mathcal{F}_{d\,\mathrm{int}}^{\mu \nu }=F_{d\,\mathrm{int}}^{\mu \nu }-%
\frac{g}{m}\epsilon _{dbc}(F_{b}^{\mu \nu }\phi _{c})_{\mathrm{int}}+ &&%
\frac{g}{m}(F_{d}^{\mu \nu }H)_{\mathrm{int}}  \nonumber \\
-\frac{g^{2}}{4m^{2}}[\delta _{da}\delta _{bc}-(\delta _{dc}\delta
_{ba}+\delta _{db}\delta _{ca})] &(:&F_{a}^{\mu \nu }\phi _{b}\phi _{c}:)_{%
\mathrm{int}}  \nonumber \\
-\frac{g^{2}}{2m^{2}}\epsilon _{dbc}(F_{b}^{\mu \nu }\phi _{c}H)_{\mathrm{int%
}}+\frac{g^{2}}{4m^{2}} &(:&F_{d}^{\mu \nu }H^{2}:)_{\mathrm{int}}+\mathcal{O%
}(g^{3}).  \label{resultF}
\end{eqnarray}

To get a better understanding of the latter result we identify these
physical fields $\mathcal{F}_{d\,\mathrm{int}}^{\mu \nu }$ (\ref{resultF})
as gauge invariant fields in the framework of spontaneous symmetry breaking
of the $SU(2)$ gauge symmetry. In this semiclassical picture the scalar
fields $\phi _{a}$ (\ref{2.35}) and $H$ (\ref{2.37}) form two $SU(2)$
doublets 
\begin{equation}
\Phi =\frac{1}{\sqrt{2}}\left( 
\begin{array}{c}
\phi _{2}+i\phi _{1} \\ 
v+H-i\phi _{3}
\end{array}
\right)  \label{Phi}
\end{equation}
and 
\begin{equation}
\tilde{\Phi}=\frac{1}{\sqrt{2}}\left( 
\begin{array}{c}
v+H+i\phi _{3} \\ 
-\phi _{2}+i\phi _{1}
\end{array}
\right) ,  \label{Phitilde}
\end{equation}
where $v$ is the vacuum expectation value of the original (i.e. non-shifted)
field $\tilde{H}=v+H$. $v$ is proportional to the gauge boson mass 
\begin{equation}
m=\frac{gv}{2}.  \label{mv}
\end{equation}
The composite fields 
\begin{equation}
\mathcal{G}_{3}^{\mu \nu }:=\tilde{\Phi}^{\ast }F^{\mu \nu }\tilde{\Phi}%
=-\Phi ^{\ast }F^{\mu \nu }\Phi =-\sum_{a=1,2,3}F_{a}^{\mu \nu }\Phi ^{\ast
}\sigma _{a}\Phi  \label{G3}
\end{equation}
($\sigma _{a}$ are the Pauli matrices) and 
\begin{equation}
\mathcal{G}_{1}^{\mu \nu }:=\frac{1}{2}(\tilde{\Phi}^{\ast }F^{\mu \nu }\Phi
+\Phi ^{\ast }F^{\mu \nu }\tilde{\Phi}),\quad \quad \mathcal{G}_{2}^{\mu \nu
}:=\frac{i}{2}(\tilde{\Phi}^{\ast }F^{\mu \nu }\Phi -\Phi ^{\ast }F^{\mu \nu
}\tilde{\Phi})  \label{G1,2}
\end{equation}
are $SU(2)$ gauge invariant, or equivalently they are invariant with respect
to the classical\footnote{%
To avoid problems of defining products of interacting fields (this can be
done by means of (\ref{int})) and the BRST-transformation thereof, we only
consider \textit{classical} fields here.} BRST-transformation 
\begin{equation}
s(F_{a}^{\mu \nu })=ig\epsilon _{abc}F_{b}^{\mu \nu }u_{c},\quad s(\phi
_{a})=imu_{a}+\frac{ig}{2}(u_{a}H+\epsilon _{abc}\phi _{b}u_{c}),\quad s(H)=-%
\frac{ig}{2}u_{a}\phi _{a}.  \label{BRS}
\end{equation}
By multiplying out the matrices in $\mathcal{G}_{d}^{\mu \nu }$ (\ref{G3}-%
\ref{G1,2}) we find that (with a suitable normalization) the corresponding
quantum fields are proportional to $\mathcal{F}_{d\,\mathrm{int}}^{\mu \nu }$%
, so far as we have determined the constant power series $t_{dab},...,\tilde{%
z}_{da}$ (\ref{ansatzF}), i.e. 
\begin{equation}
\mathcal{G}_{d\,\mathrm{int}}^{\mu \nu }(x;W)=\frac{v^{2}}{2}\mathcal{F}_{d\,%
\mathrm{int}}^{\mu \nu }(x;W)+\mathcal{O}(g^{3}).  \label{G=F}
\end{equation}
\vskip0.3cm Finally some more remarks on the validity of the
Schwinger-Swieca screening property are in order. Since the physical field
strength is an operator in Hilbert space (positive definite metric) which
fulfills a Maxwell type equation and since the particle spectrum has an
isolated mass hyperboloid for the vector meson, the theorem of Swieca \cite
{Sw} is applicable and hence the charge defined in terms of the large
surface integral over the physical field strength vanishes. This charge
screening is more interesting if additional spinor matter is present.

\section{\protect\bigskip Renormalizability for s$\geq 1$ without Ghosts?}

We have seen that the cohomological representation of the Wigner theory for
vector mesons achieves the magic trick of rescuing renormalizability,
whereas the naive application of the standard causal perturbation theory
based on interactions $W_{phys}$ in terms of local Wick polynomials in
physical (Wigner) field coordinates, which inevitably leads to $%
dimW_{phys}\geq 5$, fails on the count of renormalizability\footnote{%
This is a well-known limitation in Weinberg's program of using Wigner's
particle theory as a starting point for perturbative renormalization for
massive vector mesons and higher spin particles.}. The trick employs ghosts
in intermediate steps in a way  which is reminiscent of a catalyzer in
chemistry. This is to say a physical problem of perturbatively coupled
massive spin =1  interactions, which a priori has nothing to do with ghosts,
had to be cohomologically extended, because that was apparently the only way
to reconcile the standard perturbative machinery (of deformation of free
theories by $W^{\prime }s$) with the short distance renormalizability
requirement. But at the end, after the cohomological descend, one obtains
local physical vector meson fields and a physical $S$-matrix (both
renormalizable) within a physical Fock space of massive spin one particles,
together with new physical degrees of freedom. This is a physical result which
totally conceals the intermediate presence of ghosts. Although the final
result agrees formally with (the gauge invariant part of) renormalized gauge
theory, the underlying physical idea and the words used to describe it are
quite different. Instead of the Higgs mechanism, which generates the s=1
mass through an additionally introduced (by hand) scalar field with
non-vanishing vacuum expectation value, and which has no visible intrinsic
(gauge-invariant) meaning, the vector meson mass in the present approach is
directly linked with the aforementioned Schwinger-Swieca screening
mechanism. The latter, together with the renormalizability requirement,
demands the presence of additional degrees of freedom which we realized as
scalar particles and identified with the Higgs field without its vacuum
condensate. 

Although we have neither demonstrated uniqueness of the ghosts nor of the
new physical degree of freedom, we believe, that as already mentioned in the
third section, our minimal solution is unique in the sense that any other
solution involving higher spin ghost and induced physical objects always
contains our minimal solution (plus possible additional couplings with
coupling parameters which may be set to zero). 

Between the two formulations for interacting vector mesons, once as the
quantized version of classical gauge theories with Higgs mechanism and on
the other hand the ``ghost catalyzer'' to implement perturbative
renormalization we prefer the latter because it underlines the preliminary
aspect of our understanding of higher spin interactions more clearly. In
this sense it upholds the Bohr-Heisenberg maxim that even in cases of
discoveries via non-observable and formal constructs, one should always aim
for a reduction to observable concepts in order to obtain a conceptual
profound understanding.

The present formulation also gives a more detailed picture of physical
fields (cf. (\ref{resultF})). 
Whereas their locality and spectral properties are as in a
renormalizable lower spin s\TEXTsymbol{<}1 coupling, their operator
dimension generally does not have the form of logarithmic corrections on
canonical values of Lagrangian fields. Rather they involve higher
Wick-polynomials up to operator dimension 4 i.e. are composites in the
unphysical fields which interpolate ``elementary'' (in the sense of a
perturbative description only) physical particles. This is the consequence
of the fact that the cohomological descend renders the standard formulas in
terms of retarded $R$-products invalid. Instead of Lagrangian interpolating
fields with their canonical dimensions and logarithmic radiative corrections
we have to work with the physical composites for the interpolation of the
physical particles. The $Q$-invariant linear combinations which represent the
physical fields within the extended Wick formalism have no intrinsic meaning
or in other words the present formalism has not supplied us with a formalism
which leads to the physical vacuum expectation values in a more direct 
and less
ad hoc manner. Whereas symmetries of a free field theory can be used for the
selection of a natural invariant subalgebra, the selection of a $Q$-invariant
subalgebra of Wick-polynomials remains a rather unnatural procedure. Whereas
an observable subalgebra resulting as the fixpoint algebra of an internal
symmetry contains all the structure which via the superselection theory
allows an intrinsic reconstruction of the algebra of fields, there is no
intrinsic natural way to reconstruct the ghost extended algebra from the
physical algebra. It also indicates that even though the final physical
degrees of freedom are local and can be described in terms of covariant
point-like fields, the formalism deviates in ``some way'' from the standard
local behavior in that the perturbation cannot be interpreted as a
deformation on the original local class (Borchers class) of local physical
fields but rather appears as a subclass of a  deformed unphysical extended
theory with artificial looking rules concerning its position within the
extended local class. In this way the violation of the Bohr-Heisenberg maxim
about observables in the present description of higher spin interaction
becomes most evident. Before we present some ideas about an alternative
approach avoiding non-observable aspects, it is helpful to recall how the
Wigner theory deals with the vector potential in the case of zero mass i.e.
for photons. 

On a very formal level (ignoring regularisations of line integrals) we could
in fact also have obtained the lowering of the operator dimension dim$A$
from two to one without any cohomological extension by allowing non-point-like
and non-covariant vector potentials which are interpolating the same particles
as the covariant ones. Although on-shell objects as the $S$-matrix of the
final QFT are unaffected, as long as those fields remain almost local in the
sense of \cite{Ha}, it is not known how to deal with such objects in a
causal approach based on interaction polynomials $W$ and transition
operators $S(g)$.

It is interesting to note that the Wigner theory of free photons does not
allow the introduction of any local covariant vector potential, thus raising
the suspicion that massless spin one interactions formulated in a physical
Hilbert space lead to some sort of clash between renormalizability (which
requires vector potentials of $dimA_{\mu }=1)$ and the locality and
covariance of interaction densities W. This problem with vector potentials
comes from the non-compact structure of the little group (stability group) of
a light-like vector say $p_{R}=(1,0,0,1)$, which is the twofold covering of
the Euclidean group $E(2)$ (and is denoted by $\tilde{E}(2)$). Whereas the
rotation has the interpretation as a helicity rotation (rotation around the
third axis), the ``translations'' are Lorentz-transformations which tilt the
t-z wedge, leaving its upper light-like vector unchanged. As far as the two
transversal coordinates are concerned they behave like 2-parametric Galilei
velocity transformations (i.e. ``light cone translations'' without the
energy positivity property) with the two longitudinal light cone
translations playing the role of the Hamiltonian resp. the central mass in
the quantum mechanical representation theory of the Galilei group. The
embedding of $\tilde{E}(2)$ into $SL(2,\mathbf{C})$ for the above choice of
reference vector is 
\begin{equation}
\alpha (\rho ,\theta )=\left( 
\begin{array}{ll}
e^{i\frac{1}{2}\theta } & \rho  \\ 
0 & e^{-i\frac{1}{2}\theta }
\end{array}
\right) ,\,\,p_{R}\sim \left( 
\begin{array}{ll}
2 & 0 \\ 
0 & 0
\end{array}
\right) ,
\end{equation}
where $\theta $ is the angle for the rotations around the 3-axis and $\rho
=\rho _{1}+i\rho _{2}$ parameterizes the Euclidean translations by the
vector $(\rho _{1},\rho _{2})$. The unitary representation theory of this
non-compact group is somewhat more complicated than that of $SU(2)$. But it
is obvious that the representations fall into two classes; the ``neutrino-%
photon'' class with $U(\alpha (\rho ,0))=1$ i.e. trivial representation of
the Euclidean translations, and the remaining faithful ``continuous spin''
(infinite dimensional) representations with $U(\alpha (\rho ,0))\neq 1$. A
more detailed analysis shows that the latter lacks the strong localization
requirements which one must impose on those positive energy representations
which are used for the description of particles.\footnote{%
For the non-faithful representation ($m=0,h=$ semi-integer) ($U(\alpha (\rho
,0))\neq 1$) one finds that the infinite set of intertwiners in (\ref{fields}%
) is restricted to $u^{(n_{+},n_{-})}$ with $n_{-}=n_{+}\pm h$ \cite{We}.}
Hence the Wigner theory does not allow to describe photon operators in terms
of covariant vector fields. On the other hand a covariant field strength $%
F_{\mu \nu }$ has the following intertwiner representation in terms of the
Wigner annihilation/creation operators for circular polarized photons $%
a_{\pm }^{\#}(k)$ 
\begin{eqnarray}
F_{\mu \nu }(x) &=&\frac{1}{\left( 2\pi \right) ^{\frac{3}{2}}}\int \left\{
e^{-ikx}\sum_{\pm }u_{\mu \nu }^{(\pm )}(k)a_{\pm }(k)+h.c.\right\} \frac{%
d^{3}k}{2\left| k\right| } \\
u_{\mu \nu }^{(\pm )}(k) &\simeq &k_{\mu }e_{\nu }^{\left( \pm \right)
}(k)-k_{\nu }e_{\mu }^{\left( \pm \right) }(k)  \label{cirF}
\end{eqnarray}
Here $e_{\mu }^{\left( \pm \right) }(k)$ are the polarization vectors which
are obtained by application of the following Lorentz transformation to the
standard reference vectors $\frac{1}{\sqrt{2}}(0,\pm 1,i,0)$ (which are $%
\perp (1,0,0,1)$): a rotation of the $z$-axis into the momentum direction $%
\vec{n}\equiv \frac{\vec{k}}{\omega }$ (fixed uniquely by the standard
prescription in terms of two Euler angles) and a subsequent Lorentz-boost
along this direction which transforms $(1,\vec{n})$ into $k=\omega (1,\vec{n}%
)$. It is these vectors that do not behave covariant under those
Lorentz-transformations which involve the above ``little group
translations'' but rather produce an affine transformation law 
\begin{equation}
G(\rho )e^{(\lambda )}(p_{R})=e^{(\lambda )}(p_{R})+\left\{ 
\begin{array}{c}
-\frac{1}{2}\left( \bar{\rho},0,0,\bar{\rho}\right) ,\,\,\,\,\,\,\lambda =+
\\ 
+\frac{1}{2}\left( \rho ,0,0,\rho \right) ,\,\,\,\,\,\,\,\lambda =-
\end{array}
\right. \,\,\,\,\,\,\,\,\,\,\,\,\,\,\,\,\,\,\rho =\rho _{1}+i\rho _{2},
\end{equation}
($G(\rho )$ is the Minkowski space representation of the Euclidean
translations by $(\rho _{1},\rho _{2})$) whereas under $x$-$y$ rotations the 
$e^{(\lambda )}$ picks up the standard Wigner phase factor. The polarization
vectors do not behave as 4-vectors since they are not invariant under the
Euclidean translations in $\tilde{E}(2)$, as one would have expected for a
(non-existing!) bona fide intertwiner from the $(0,h=1)$ Wigner
representation to the $D^{\left[ \frac{1}{2},\frac{1}{2}\right] }$ covariant
representation. Rather the intertwiner only has Lorentz-covariance 
up to additive
gauge transformations i.e. up to affine longitudinal terms. For general
Lorentz transformations this affine law reads: 
\begin{equation}
(U(\Lambda )e)_{\mu }(k)=\Lambda _{\mu }^{\nu }e_{\nu }(\Lambda
^{-1}k)+k_{\mu }H(\Lambda ,k)
\end{equation}
This peculiar manifestation of the $\left( 0,h=1\right) $ little group $%
\tilde{E}(2)$ is the cause for the appearance of the local gauge issue in
local quantum physics. Unfortunately this quantum origin is somewhat hidden
in the quantization approach, where it remains invisible behind the
geometrical interpretation in terms of fibre bundles. In terms of the
potential in the physical Fock space we have 
\begin{equation}
U(\Lambda )A_{\mu }(x)U^{\ast }(\Lambda )=\Lambda _{\mu }^{-1\nu }A_{\nu
}(\Lambda x)+\partial _{\mu }H
\end{equation}
where $H$ is a concrete operator involving the $a_{\pm }^{\#}$ and $e_{\pm }$
from (\ref{cirF}). Formally the string-like localized vector potential (with
localization chosen in the space-like $n$-direction) may be written as a line
integral over the field strength 
\begin{eqnarray}
&&\partial _{\mu }A_{\nu }(x)-\partial _{\nu }A_{\mu }(x)=F_{\mu \nu }(x) \\
&&A_{\nu }=\frac{1}{n\cdot \partial }n^{\mu }F_{\mu \nu }\curvearrowright
\,n\cdot A=0.  \nonumber
\end{eqnarray}
The regularization needed in order to make mathematical sense out of these
string-like objects and related technical problems will not be discussed in
this qualitative presentation. The physical manifestation of the existence
of the somewhat nonlocal vector potential is the breakdown of the additivity
of the algebras generated by the field strength for non-simply connected
regions \cite{Ha}. The natural localization regions for vector potentials
are space-like cones or wedges which contain such cones. On the other hand
the causal perturbation theory (and any other formulation using Lagrangian
quantization) requires point-like fields.

Comparing this with the massive case, one notes that the problematic aspects
of the use of vector potentials in the local description of s=1 show up in
the quantum properties of free Wigner photons before the renormalization
procedure, whereas for massive vector mesons it appears only in facing the
issue of renormalizability (dimensional counting) of interactions. What is
in common to both cases is the fact that no existing formulation of
renormalized perturbation theory is capable to deal with dim$A_{\mu }=1$
nonlocal vector potentials within the framework of causal Wick-polynomials $W.
$ Both cases can be dealt with in terms of the same remedy namely
cohomological extension by ghosts within a BRS-like formalism; however as a
result of the availability of arguments based on scattering theory, the BRS
formalism is conceptually much simpler in the massive case, whereas in the
massless theory one must use the full BRS-formalism which suffers
contribution from interactions and requires additional concepts for 
the separation of observable algebras from physical states \cite{D-F}. 

It should be clear from these remarks that in order to follow the
Bohr-Heisenberg maxim and remove the ``ghost-catalyzers'' in favor of a
ghostfree formulation, one has to go significantly outside the present
perturbative framework. A hint where to look comes from the observation that
the ghosts have been introduced to lower the operator dimension of $A_{\mu }$
from two to one, while still maintaining the (formal) locality and the
covariance of the operators. This is clearly an off-shell short distance
argument. So this observation suggests to look for an on-shell formulation.
On-shell quantities are the true S-matrix and formfactors of physical
operators $A$ between multi-particle scattering states: 
\begin{eqnarray}
&&S\left| p_{1},...p_{n}\right\rangle ^{out}=\left|
p_{1},...p_{n}\right\rangle ^{in}  \nonumber \\
&&^{out}\left\langle p_{1}^{\prime }...p_{m}^{\prime }\right| A\left|
p_{1},...p_{n}\right\rangle ^{in}\label{form}
\end{eqnarray}
Note that we defined the formfactors in such a way that the matrix elements
of $S$ themselves correspond to the formfactor of the identity operator. So
the first question is: can one compute the perturbative expansion of $S$
while staying on-shell all the time. The causal approach used before does
not fulfill this requirement, since the Bogoliubov transition operator $S(g)$
is off shell (in the sense of this paper) and only approaches the mass shell
in the adiabatic limit\footnote{%
The difference between on- and off-shell looks quite innocuous in momentum
space, however the spacetime aspects could be different. In order to
illustrate this point, just look at the massive Thirring model. The $S(g)$
resp. the field correlation functions show the full field theoretic vacuum-
and one-particle polarization structure (virtual particle structure) whereas
the $S$-matrix and the closely related PFG wedge generators 
(which are introduced below) are ``quantum
mechanical'' i.e. obey particle number conservation.}. In fact it is not
possible to formulate causality which involves products of fields directly
on-shell, rather the on-shell substitute is the notoriously elusive crossing
symmetry together with some inexorably linked subtle analytic continuation
properties. So the question is: does there exist a formulation of quantum
field theory which, different from quantization, uses such on-shell concepts
in its construction. 

Fortunately the beginning of such a new approach to QFT already exists \cite
{S-W}\cite{Schr4}. This approach can be viewed as a generalization of the
Wigner theory in the presence of interactions. It bypasses completely the
use of point-like fields and aims directly at the wedge algebras and the
associated net of smaller localized algebras obtained by intersecting wedge
algebras. 

It is deeply satisfying that the modular theory used in this construction
directly converts the Wigner one particle theory into the interaction free
nets of algebras without using fields and their equivalence classes in
intermediate steps. The generalization to interacting models 
is given by the modular
interpretation of the S-matrix as a relative modular invariant which
``measures'' the interaction for wedge-localized algebras. The naturalness
of the wedge region appeared for the first time long ago in Unruh's
Minkowski space analogue of the Hawking thermal aspects of black holes.
Indeed, its use for a generalization of the Wigner approach in the presence
of interactions leads rather swiftly to a profound understanding of two
aspects: the relation of the thermal aspect of modular (wedge) localization
with the aforementioned elusive crossing symmetry of particle physics, and
the existence of generators of the wedge algebra which, applied to the vacuum,
create one particle states free of particle-antiparticle polarization
clouds, despite the presence of interactions. \ The crossing is known to be
closely related to the TCP theorem; in fact it is a kind of TCP property for
individual particle in formfactors whereby the particle in an incoming
configuration is flipped into an (analytically continued) outgoing
anti-particle. The existence of (vacuum) \textbf{p}olarization \textbf{f}ree 
\textbf{g}enerators (abbreviated ``PFG'') for wedge localized algebras on
the other hand is a new discovery. Their correlation functions can be
expressed in terms of \ ``nested'' products of S-matrices \cite{Schr4}. A
closer look has revealed that in the special case of d=1+1 purely elastic
S-matrices (``factorizing models'' \cite{Ka}\cite{Smi}) the PFG's in fact
obey the Zamolodchikov-Faddeev algebra, and that the latter receives in turn
for the first time a physical spacetime interpretation which transcends its
formal use \cite{Schr1}\cite{Schr2}\cite{Schr3}\cite{Schr4}. 

The potential relevance of these remarks for a ghostfree formulation of
higher spin interactions becomes clearer if one notices that the reasons for
their introduction is that the point-like interaction densities $W_{phys}$ in
terms of physical higher spin fields have a short distance behavior beyond
the one allowed by the formal renormalizability criterion. If we could avoid
such off-shell objects as point-like fields and their correlation functions
in favor of formfactors of localized operators (\ref{form}) then the
existence of the theory would not be threatened by the short distance
behavior in the construction of the theory. Rather it would come out from the 
definition of formfactors of products of such operators by summing over the
complete set of intermediate particle states. The first step in such a new
construction program namely the construction of the PFG's belonging to the
model is purely on-shell and only involves physical particles. The
correlation functions of PFG's are on-shell objects, in the old fashioned
language they contain a ``natural cutoff'' as a result of their somewhat
de-localized nature, but unlike the brutal force cutoffs in the Lagrangian
approach, these objects coexist in the same Hilbert space together with
sharper localized operators which belong to intersections of wedge algebras.
The non-triviality check in this approach, which is expected to replace the
good short distance behavior in the standard renormalization theory, is the
non-triviality of algebras localized in double cones obtained by intersecting
wedges
\begin{equation}
\mathcal{A}(\mathcal{O})=\cap _{\mathcal{O}\subset W}\mathcal{A}(W)
\end{equation}
It may not be widespread known, but it is nevertheless true that all
physical informations are contained in local algebras\footnote{%
This independence of the physical content from field coordinatizations is
the main reason for the interest in the algebraic approach
(i.e. local quantum physics).} and that there
is no necessity from a conceptual point of view to use point-like field
coordinates apart from certain distinguished point-like currents related to
symmetries and possibly associated order/disorder fields.

The first step, namely the construction of the on-shell PFG's is reminiscent
of attempts in the 60$^{ies}$ to construct unitary crossing symmetric
S-matrices. This attempt, even in its limited perturbation version, was
ill-fated and did not lead to tangible results. It remained as a very
fundamental problem in particle physics closely related to the inverse
problem i.e. the question to what extend a physically admissible (on-shell)
S-matrix determines uniquely an (off-shell) algebraic net. The modular
approach to wedge algebras and the concept of PFG's incorporates and
enriches this old S-matrix program, so that it becomes the first step in a
construction of local nets. The hope that in this new version it becomes at
least susceptible to a kind of on-shell perturbation theory rests on the new
field theoretic setting together with rich new concepts which were absent in
the days of the old bootstrap program. In fact the success of this program
for d=1+1 factorizing models \cite{Ka}\cite{Smi} (in which case such an
S-matrix which avoids the standard construction exists even in a summed up
non-perturbative form) may be taken as an encouragement for a future
exploration of this new approach to QFT.

One expects that this new approach reproduces the perturbative results of
the standard renormalization theory wherever the latter is applicable. The
innovative power of the new approach is expected to show up in situations
where one has reasons to believe that Lagrangian quantization does not
provide physical field coordinates of sufficient low operator dimension
related to good short distance properties needed for the W's in the causal
approach. There is no a priori reason why the Lagrangian fields should be
the ones with the lowest short distance dimensions in the local equivalence
class of all local fields of one model. In the case of spin=1 one can still
overcome such obstacles against the standard formulations by the ad hoc
ghost trick which is suggested by the quantization of gauge theories. But 
could there also be (e.g. higher spin) cases with a finite number of coupling
parameters which are declared ``non-renormalizable'' in the standard
approach, even though they have a well-defined perturbation theory in the
new sense? 

\section{Future Perspectives}

The history of gauge theory and its connection with higher spin
renormalizable QFT is one of the theoretically most  fascinating and
experimentally most significant developments in the 20$^{th}$ century.
Contrary to a widely held opinion  this is still an unfinished story with
possible future surprises. 

The usefulness of the role of the ``gauge principle'' as a selection
principle in the classical setting and its attractive geometric appeal in
the semiclassical realm of quantum matter coupled to external fields (minimal
electromagnetic-coupling) is universally recognized. Acknowledging its
important role in the discovery of the renormalization of interactions
between massive vector mesons, we nevertheless tried to argue in this paper
that further progress in this area requires new ideas beyond those of
standard gauge theory, which are of a more algebraic 
(or local quantum physical)
than differential geometric kind. As a compromise between the standard
approach and a future ghostfree framework we presented a formulation of
interactions of massive vector mesons using ghosts which extend the Wigner
one-particle states of the vector meson. In this formulation the 
necessity of the presence of further
Higgs like physical degrees of freedom is most easily recognizable, even
though they have been stripped of their role of ``fattening'' the
photons/gluons into massive vector mesons via Higgs condensates 
since their mass was
assumed  non-vanishing from the outset. It is interesting to note that even
without this role their presence is nevertheless necessary for perturbative
consistency and the Schwinger-Swieca charge screening mechanism. Although we
hope that this formulation will only be intermediary and replaced eventually
by a ghostfree modular approach (the remarks in the previous section), it is
in our view already with all its shortcomings better suitable than the
standard gauge approach to highlight those points which suggest the
necessity of a radically different formulation. 

The viewpoint in this paper is in some ways opposite to the standard one.
Whereas standardly the massless case is used as the basic reference and the
mass of vector mesons is attributed to the Higgs mechanism in gauge theories
which follows more the geometric logic of classical field theory, we have
favored a particle viewpoint which emphasizes the physical particle 
interpretation and leaves the more difficult off-shell infrared
rearrangements in the massless limit to future more detailed investigations.
Whether a completely ghostfree formulation (which fully explains the
gauge-restricted structure of interacting higher spin objects as
consequences of the same basic local quantum physical properties which hold
for low spins) is feasible, only future can tell.     

We also mentioned various historical precedents to our point of view.
In this context it is interesting to add a quotation from a paper of P.W
Anderson \cite{An}. About the possible non-intrinsic aspect of the Higgs
mechanism for massive photons in comparison to plasmons he writes: ``How,
then, if we were confined to a plasma as we are to the vacuum and could only
measure renormalized quantities, might we try to determine whether, before
turning on the effects of electromagnetic interaction, $A$ (the
vector potential) had been a massless gauge field and... ? As far as we can
see this is not possible.`` The lack of intrinsic physical meaning of Higgs
condensates was also the main issue of the well-known Elitzur's theorem.     

If by our observations on massive vector mesons we succeed to re-direct some
of the attention which has been given  to geometric/mathematical aspects
towards the many open and interesting conceptual local quantum problems, the
time it took for writing this paper was well spent.

\textbf{Acknowledgments:}

The authors thank Klaus Fredenhagen for helpful suggestions and one of us
(M. D.) is indebted to G\"{u}nter Scharf for several fruitful discussions.

\section{Appendix A: The $S$-matrix to first order}

First we give an incomplete proof of the following statement (which is of
general interest):

\textbf{Conjecture:} Let $V(x)$ be an arbitrary Wick polynomial with
derivatives (i.e. an arbitrary element of the Borchers class of the free
fields) and let all free fields be massive. Then the relation 
\begin{equation}
\lim_{\epsilon\to 0}\int d^4x\,V(x)g(\epsilon x)=0,\quad\quad (g\in \mathcal{%
D}(\mathbf{R}^4),\quad g(0)=1)  \label{A.1}
\end{equation}
implies that there exists another Wick polynomial (with derivatives) $%
W_1^\mu $ such that 
\begin{equation}
V=\partial_\mu W_1^\mu +V_0.  \label{A.2}
\end{equation}
Thereby $V_0$ is the sum of all Wick monomials $C:\partial^{a_1}A_{j_1}(x)
...\partial^{a_n}A_{j_n}(x):$ ($C=$const., $A_j$ is a free field with mass $%
m_j$) in $V$ for which the masses $m_{j_1},...,m_{j_n}$ are such that the
solution for $(k_1,...,k_n)\in \mathbf{R}^{4n}$ of 
\begin{equation}
k_1+...+k_n=0\quad\bigwedge\quad k_l^2=m_{j_l}^2\quad\forall l=1,...,n
\label{A.2a}
\end{equation}
contains no non-empty open subset of the manifold $k_l^2=m_{j_l}^2\quad%
\forall l=1,...,n$.

A simple example for a term in $V_0$ is $:\varphi (x)^3:$, where $(\square
+m^2)\varphi =0,\> m>0$.

\textit{Incomplete proof:}\footnote{%
Klaus Fredenhagen told us the main idea of proof. The method is strongly
influenced by the proof of lemma 2 (including appendix (b)) in \cite{Bu-F}.}
To simplify the notations we assume that the free fields are bosonic
scalars, the generalization to other types of fields is obvious. Global
factors $2\pi $ are omitted in the whole proof. We write $V$ in the form 
\begin{equation}
V(x)=\sum_{n}\sum_{j_{1},...,j_{n}}P_{j_{1}...j_{n}}(\partial
^{x_{1}},...,\partial
^{x_{n}}):A_{j_{1}}(x_{1})...A_{j_{n}}(x_{n}):|_{x_{1}=...=x_{n}=x},
\label{A.3}
\end{equation}
where the sum over $n$ is finite, the $A_{j}$'s are free fields and $%
P_{j_{1}...j_{n}}(\partial ^{x_{1}},...,\partial ^{x_{n}})$ is a polynomial
in the partial derivatives $\partial _{\mu }^{x_{l}},\>1\leq l\leq n$.
Obviously we may replace $P_{j_{1}...j_{n}}(\partial ^{x_{1}},...,\partial
^{x_{n}})$ by 
\begin{equation}
\bar{P}_{j_{1}...j_{n}}(\partial ^{x_{1}},...,\partial ^{x_{n}}) \equiv\frac{%
1}{n!} \sum_{\pi \in \mathcal{S}_n}P_{j_{\pi (1)}...j_{\pi (n)}}(\partial
^{x_{\pi (1)}},...,\partial ^{x_{\pi (n)}})  \label{A.4}
\end{equation}
in (\ref{A.3}). In terms of annihilation and creation operators the free
fields read 
\begin{equation}
A_{l}(y)=\int d^{4}p\,\delta (p^{2}-m_{l}^{2})\bar A_l(p)e^{ipy},\quad \bar
A_l(p):=[\Theta (-p^{0})a_{l}(-{\vec{p}})+ \Theta (p^{0})b_{l}^{+}({\vec{p}}%
)]  \label{A.5}
\end{equation}
($a_{l}=b_{l}$ is possible), where $[b_{l}({\vec{p}}),b_{j}^{+}({\vec{k}}%
)]=\delta _{lj}2\sqrt{{\vec{p}}^{2}+m_{l}^{2}}\delta ^{3}({\vec{p}}-{\vec{k}}%
)$ and similar for $[a_{l},a_{j}^{+}]$. Now we insert (\ref{A.3}-\ref{A.5})
into (\ref{A.1}) 
\begin{eqnarray}
0 &=&\lim_{\epsilon\to 0} \sum_{n}\int dx\,g(\epsilon x)\int
dx_{1}...dx_{n}\,\delta (x_{1}-x,...,x_{n}-x)\cdot  \nonumber \\
&\cdot &\sum_{j_{1},...,j_{n}}\bar{P}_{j_{1}...j_{n}}(\partial
^{x_{1}},...,\partial ^{x_{n}}):A_{j_{1}}(x_{1})...A_{j_{n}}(x_{n}): 
\nonumber \\
&=&\sum_{n}\int d^{4}p_{1}...d^{4}p_{n}\,\sum_{j_{1},...,j_{n}}\bar{P}%
_{j_{1}...j_{n}}(ip_{1},...,ip_{n})\delta ^{4}(p_{1}+...+p_{n})\cdot 
\nonumber \\
&\cdot &:\prod_{l=1}^{n}\delta (p_{l}^{2}-m_{j_{l}}^{2})[\Theta
(-p_{l}^{0})a_{j_{l}}(-{\vec{p}}_{l})+\Theta (p_{l}^{0})b_{j_{l}}^{+}({\vec{p%
}}_{l})]:.  \label{A.6}
\end{eqnarray}
(Here we have exchanged the order of the integrations and the limit $%
\epsilon \rightarrow 0$. This can be justified by considering matrix
elements between wave packets.) Taking improper matrix elements $%
<b_{r_{1}}^{+}({\vec{k}}_{1})...b_{r_{s}}^{+}({\vec{k}}_{s})\Omega
|...a_{r_{s+1}}^{+}(-{\vec{k}}_{s+1})...a_{r_{t}}^{+}(-{\vec{k}}_{t})\Omega
> $ of (\ref{A.6}), where $\Omega $ is the vacuum of the free fields, we
obtain\footnote{%
Usually there are several possibilities to contract the creation and
annihilation operators. All give the same contribution due to the
permutation symmetry of $\bar{P}_{r_{1}...r_{t}}$ (\ref{A.4}).} 
\begin{equation}
0=\bar{P}_{r_{1}...r_{t}}(ik_{1},...,ik_{t})\delta
^{4}(k_{1}+...+k_{t})\quad \mathrm{where}\quad k_{l}=(\pm \sqrt{{\vec{k}}%
_{l}^{2}+m_{r_{l}}^{2}},{\vec{k}}_{l})\>\>\forall l,  \label{A.7}
\end{equation}
more precisely $k_{l}^{0}>0$ for $l=1,...,s$ and $k_{l}^{0}<0$ for $%
l=s+1,...,t$. If $r_{1},...,r_{t}$ are such that the corresponding term
belongs to $(V-V_{0})$, then the condition (\ref{A.7}) restricts the
polynomial $\bar{P}_{r_{1}...r_{t}}$ in the following way. We make a Taylor
expansion with respect to $k=(k_{1}+...+k_{t})$ at $k=0$: 
\begin{equation}
\bar{P}_{r_{1}...r_{t}}(ik_{1},...,ik_{t})=\bar{P}_{r_{1}...r_{t}}(ik_{1}
,...,ik_{t-1},-i(k_1+...+k_{t-1}))+ik_{\mu }\tilde Q_{r_{1}...r_{t}}^{\mu
}(k_{1},...,k_{t})  \label{A.8}
\end{equation}
where $\tilde Q_{r_{1}...r_{t}}^{\mu }$ is also a polynomial. From (\ref{A.7}%
) we conclude (for all $(r_{1},...,r_{t})$ which appear in $(V-V_{0})$) that 
$\bar{P}_{r_{1}...r_{t}}(ik_{1},...,ik_{t-1},-i(k_1+...+k_{t-1}))$ vanishes
on a non-empty open subset of the manifold $k_i^2=m_{r_i}^2\>\forall
i=1,...,t-1$ and $(k_1+...+k_{t-1})^2=m_{r_t}^2$. We strongly presume that
this implies that there exist other polynomials $P^{(l)}(k_{1},...,k_{t-1})$ 
$(l=1,...,t)$ (for simplicity we omit the indices $r_{1},...,r_{t}$) such
that 
\begin{eqnarray}
\bar{P}(ik_{1},...,ik_{t-1},-i(k_1+...k_{t-1}))&=&\sum_{l=1}^{t-1}(k_l^2-
m_{r_l}^2)P^{(l)}(k_{1},...,k_{t-1})  \nonumber \\
&+&((k_1+...+k_{t-1})^2-m_{r_t}^2)P^{(t)}(k_{1},...,k_{t-1}).  \label{A.8a}
\end{eqnarray}
For vanishing masses this conjecture has been proved by Buchholz and
Fredenhagen (in appendix (b) of \cite{Bu-F}).\footnote{%
Unfortunately a straightforward generalization of their proof does not work.
However, for our purpose we do not need the validity of (\ref{A.8a}).} By
means of (\ref{A.8a}) we conclude 
\begin{eqnarray}
\bar{P}(ik_{1},...,ik_{t})\prod_{l=1}^{t}\delta (k_{l}^{2}-m_{r_{l}}^{2})
&=& [((k_1+...+k_{t-1})^2-k_t^2) P^{(t)}(k_{1},...,k_{t-1})  \nonumber \\
&&+ik_{\mu }\tilde Q^{\mu}(k_{1},...,k_{t})]\prod_{l=1}^{t}\delta
(k_{l}^{2}-m_{r_{l}}^{2})  \nonumber \\
=i(k_1+...+k_t)_\mu &Q^{\mu}(ik_{1},...,ik_{t})&\prod_{l=1}^{t}\delta
(k_{l}^{2}-m_{r_{l}}^{2}),  \label{A.9}
\end{eqnarray}
where $Q^{\mu}(ik_{1},...,ik_{t}):=-i(k_1+...+k_{t-1}-k_t)^\mu
P^{(t)}(k_{1},...,k_{t-1})+\tilde Q^{\mu}(k_{1},...,k_{t})$. Next we
symmetrize $Q^\mu$ according to (\ref{A.4}) and denote the result by $\bar
Q^\mu$. The equation (\ref{A.9}) holds still true if we replace $Q^\mu$ by $%
\bar Q^\mu$. Summing up we obtain 
\begin{eqnarray}
(V-V_{0})(x) &=&\sum_{n}\sum_{j_{1},...,j_{n}}\int
d^{4}p_{1}...d^{4}p_{n}\,i(p_{1}+...+p_{n})_{\mu }\bar
Q_{j_{1}...j_{n}}^{\mu }(ip_{1},...,ip_{n})\cdot  \nonumber \\
&&\cdot e^{i(p_{1}+...+p_{n})x}:\prod_{l=1}^{n}\delta
(p_{l}^{2}-m_{j_{l}}^{2})[\Theta (-p_{l}^{0})a_{j_{l}}(-{\vec{p}}%
_{l})+\Theta (p_{l}^{0})b_{j_{l}}^{+}({\vec{p}}_{l})]:  \nonumber \\
&=&\partial _{\mu }^{x}\sum_{n}\sum_{j_{1},...,j_{n}}\bar Q_{j_{1}...j_{n}}
^{\mu}(\partial ^{x_{1}},...,\partial
^{x_{n}}):A_{j_{1}}(x_{1})...A_{j_{n}}(x_{n}):|_{x_{1}=...=x_{n}=x} 
\nonumber
\end{eqnarray}
(we set $\bar Q_{r_{1}...r_{t}}^{\mu }:=0$ if $(r_{1},...,r_{t})$ belongs to 
$V_0$) which is the assertion (\ref{A.2}).

We do not need the general result (\ref{A.1}-\ref{A.2a}), we only deal with
very special situations. Physical consistency (\ref{2.31}) to first order
means 
\begin{equation}
\lim_{\epsilon\to 0}\int d^4x\,g(\epsilon x)[Q,W(x)]\vert_{\mathrm{ker}\,Q}
=0.  \label{A.11}
\end{equation}
Let us first consider the \textit{quadrilinear} terms in $V(x):=[Q,W(x)]$,
i.e. $[Q,W(x)^{(4)}]$, where $W^{(4)}\sim :AAAA:,\,\sim :AAu\tilde u:,
\,\sim :u\tilde u u\tilde u:,\,\sim:\phi\phi\phi\phi:,\,\sim:\phi\phi AA:,
\,\sim :\phi\phi u\tilde u:,\,\sim :\phi\phi H^2:,\,\sim :AAH^2:,\,\sim
:u\tilde uH^2:,\,\sim :H^4:$ (the upper index $(4)$ refers to the
quadrilinear terms). They satisfy the kinematical equations (\ref{A.2a}) for
a sufficient big set of momenta, i.e. they belong to $(V-V_0)$.\footnote{%
For a free field $\varphi$ note that $[Q,\varphi (x)]$ satisfies the same
field equation as $\varphi$ and hence has the same mass.} The polynomials $%
P_{j_{1}...j_{n}}$ (\ref{A.3}) (and hence also $\bar P$ (\ref{A.4})) are of
the form 
\begin{equation}
\bar P(ik_1,...,ik_4)=c_0\quad\quad\mathrm{or}\quad\quad \bar
P^\mu(ik_1,...,ik_4)=\sum_{i=1}^4c_ik_i^\mu,  \nonumber
\end{equation}
where $c_0$ and $c_i$ are constants. Taking matrix elements of (\ref{A.11})
with improper two-particle states, where the incoming state is an element of 
$\mathrm{ker}\,Q$, we obtain (\ref{A.7}), which implies $\bar P=0$ or $\bar
P^\mu =c(k_1^\mu+...+k^\mu_4)$ ($c=$constant). Hence (\ref{A.11}) implies
that there exists a Wick polynomial $W_1^{(4)\nu}$ such that 
\begin{equation}
[Q,W(x)^{(4)}]=i\partial_\nu W_1^{(4)\nu}(x).  \nonumber
\end{equation}
This is the Q-divergence condition to first order (\ref{2.29}). In addition $%
W^{(4)}$ must satisfy the requirements (a)-(e) listed in section 2. The only
solution is $\sim :H^4:$ \cite{G}, but such a term will be excluded by
physical consistency (\ref{2.31}) to second order.

A \textit{trilinear} term in $W(x)$ (with masses $m_1,\,m_2$ and $m_3$)
fulfills the kinematical equations (\ref{A.2a}) for a sufficient big set iff
there exists a permutation $\pi\in \mathcal{S}_3$ such that 
\begin{equation}
m_{\pi (3)}>m_{\pi (1)}+m_{\pi (2)}.  \label{A.13}
\end{equation}
But physical consistency to higher orders will require $m_1=m_2=m_3=:m$
apart from the $H$-couplings ($m_H>0$ will not be restricted; see sect. 2).
Neglecting this exception we conclude from (\ref{A.6}) that the trilinear
part $S_1^{(3)}$ of the $S$-matrix to first order vanishes in the adiabatic
limit, i.e. 
\begin{equation}
S^{(3)}_1=\lim_{\epsilon\to 0}\int d^4x\,g(\epsilon x)W^{(3)}(x)=0,
\label{A.13a}
\end{equation}
and $[Q,S^{(3)}_1]=0$ (\ref{2.31}) is trivially satisfied. To get
restrictions on $W^{(3)}$ we must consider physical consistency (\ref{2.31})
to orders $n\geq 2$, especially for second order tree diagrams.

\section{Appendix B: Determination of the parameters in $W$ and $\mathcal{F}%
_{\mathrm{int}}$ by requiring $[Q,\mathcal{F}_{\mathrm{int}} (x;W)]=0$}

The calculation is lengthy, hence we concentrate on the essential points. We
determined the parameters in the Ansatz for $W$ (\ref{ansatzW}) and $\mathcal{%
F}_{\mathrm{int}}$ (\ref{ansatzF}) by inserting these expressions into the
physicality requirement (\ref{QF'}) to lowest perturbative orders. Once a
parameter is fixed, its value will be used in the following calculations
without mentioning it.

- To zeroth order the condition is trivially fulfilled due to\footnote{%
If we would not have made the assumption $c^{(0)}_k=0$ (i), we would find
here that most of the zeroth order coefficients must vanish. It would remain 
$\mathcal{F}^{\mu\nu\,(0)}_d=F^{\mu\nu}_d+x^{(0)}_{da}F^{\mu\nu}_aH
+z^{(0)}_{da}:F^{\mu\nu}_aH^2:$ only. $x^{(0)}$ and $z^{(0)}$ are then
forced to vanish by higher orders of (\ref{QF'}). But this complicates the
calculation a lot and the assumption $c^{(0)}_k=0$ is reasonable. So we do
not do this effort here.} 
\begin{equation}
[Q,F^{\mu\nu}_d]=0.  \label{QF0}
\end{equation}

- To first order the equation 
\begin{equation}
\lbrack Q,\mathcal{F}_{d}^{(1)}(x)]+i\int dx_{1}\,[Q,R(W(x_{1});F_{d}(x))]=0
\label{QF1}
\end{equation}
is required. The second term contains tree diagrams only. By means of (\ref
{retprod}) they can be written in the form 
\begin{eqnarray}
R(W(x_{1});F_{d}^{\mu \nu }(x)) &=&\Theta (x^{0}-x_{1}^{0})[F_{d}^{\mu \nu
}(x),W(x_{1})]  \nonumber \\
+\delta (x-x_{1})iC\frac{1}{2}(f_{bcd}-f_{cbd}) &:&A_{b}^{\mu
}(x_{1})A_{c}^{\nu }(x_{1}):,  \label{RWF}
\end{eqnarray}
where $C$ is an arbitrary constant. The terms $\sim \Theta (x^{0}-x_{1}^{0})$
have propagators $\Delta ^{\mathrm{ret}}(x-x_{1}),\,\partial \Delta ^{%
\mathrm{ret}}(x-x_{1}),\,\partial \partial \Delta ^{\mathrm{ret}}(x-x_{1}),$
where $\Delta ^{\mathrm{ret}}$ is the retarded fundamental solution of the
Klein Gordon equation, $\mathrm{supp}\,\Delta ^{\mathrm{ret}}\subset \bar{V}%
^{+}$. The term $\sim C\delta (x-x_{1})$ is due to the non-uniqueness of the
extension of $T(W(x_{1})F_{d}^{\mu \nu }(x))$ to the diagonal $x_{1}=x_{2}$.
This is similar to the tree normalization terms (\ref{2.49}). The tensor $%
f_{bcd}$ which is the same as in the Ansatz for $W$\footnote{%
We use here an additional normalization condition, namely that the term $%
\sim C\delta (x-x_{1})\break :AA:$ has the same color tensor as the
corresponding non-local term (i.e. the term in $\Theta
(x^{0}-x_{1}^{0})[F^{\mu \nu }(x),W(x_{1})]$ with the same external legs).}
is antisymmetrized according to the normalization condition $\mathcal{F}_{d\,%
\mathrm{int}}^{\mu \nu }=-\mathcal{F}_{d\,\mathrm{int}}^{\nu \mu }$ (iv).
Now let $\psi _{1},\,\psi _{2}$ be arbitrary free fields and $y_{1},y_{2}\in 
\mathbf{R}^{4},\>(y_{j}-x)^{2}<0\>\>(j=1,2)$. By commuting (\ref{QF1}) with $%
\psi _{1}(y_{1})$ and $\psi _{2}(y_{2})$ we obtain 
\begin{equation}
0=\int dx_{1}\,[[[Q,[F_{d}(x),W(x_{1})]],\psi _{1}(y_{1})],\psi
_{2}(y_{2})]\Theta (x^{0}-x_{1}^{0}).  \label{QF1'}
\end{equation}
We conclude that for $x_{1}\in (x+\bar{V}^{-})\setminus \{x\}$ the
expression $[Q,[F_{d}(x),W(x_{1})]]$ must be a divergence of a local
operator with respect to $x_{1}$,\footnote{%
Note that every operator valued distribution $G(x_{1},...,x_{n}),\>(x_{j}\in 
\mathbf{R}^{4})$ can be written as divergence of a \textit{non-local}
operator, e.g. $G(x_{1},...,x_{n})=\partial _{\nu }^{x_{1}}\int dy\,\partial
^{\nu }D^{...}(x_{1}-y)G(y,x_{2},...,x_{n})$, where $D^{...}$ is a
fundamental solution of the wave equation.} i.e. there must exist a
two-point distribution $R_{1d}^{\tau }(x_{1},x)$ with 
\begin{equation}
\Theta (x^{0}-x_{1}^{0})[Q,[F_{d}(x),W(x_{1})]]=\partial _{\tau
}^{x_{1}}R_{1d}^{\tau }(x_{1},x)+\delta (x-x_{1})P_{d}(x),  \label{QF1''}
\end{equation}
where $P_{d}(x)$ is a Wick polynomial (with constant coefficients) in free
fields (possibly with derivative). In contrast to the adiabatic limit of $W$
(\ref{A.1}) there is no kinematical restriction (in the sense of (\ref{A.2a}%
)) here, because one of the free field operators in $W(x_{1})$ is contracted
with $F(x)$ and, hence, the corresponding momentum must not be on-shell. We
have not worked out a general proof (in the style of appendix A) of the step
from (\ref{QF1'}) to (\ref{QF1''}). But the explicit calculations show that
this conclusion is correct.\footnote{%
This remark concerns also the analogous (more complicated) steps from (\ref
{Qtree2}) to (\ref{QF2'}) and from (\ref{QF3'}) to (\ref{QF3''}) in second
and third order.} Inserting (\ref{QF1''}) and (\ref{RWF}) into (\ref{QF1})
it results 
\begin{equation}
\lbrack Q,\mathcal{F}_{d}^{\mu \nu \,(1)}(x)]+iP_{d}^{\mu \nu }(x)-C\frac{1}{%
2}(f_{bcd}-f_{cbd})[Q,:A_{b}^{\mu }(x)A_{c}^{\nu }(x):]=0,  \label{QF1lo}
\end{equation}
which means that the ''local terms''\footnote{%
We set quotation marks because the splitting of a distribution in a local
and a non-local part is non-unique.} must satisfy the condition (\ref{QF1})
separately. Later when we shall know more about $W$ (which determines $P_{d}$
by (\ref{QF1''})), we will compute most of the coefficients in $\mathcal{F}%
_{d}^{(1)}$ (\ref{ansatzF}) and the normalization constant $C$ from (\ref
{QF1lo}). Inserting (\ref{QF0}) into (\ref{QF1''}) we find that $W$ must
fulfill 
\begin{equation}
\lbrack F_{d}(x),[Q,W(x_{1})]]=\partial _{\tau }^{x_{1}}R_{1d}^{\tau
}(x_{1},x)\quad \mathrm{for}\quad x_{1}\in (x+\bar{V}^{-})\setminus \{x\}.
\label{FQW}
\end{equation}
Obviously this condition is truly weaker than $[Q,W]=$(divergence of a Wick
polynomial) (\ref{2.29}). But the higher orders of the physicality
requirement (\ref{QF'}) yield more information about $W$.

- To second order (\ref{QF'}) reads 
\begin{eqnarray}
\lbrack Q,\mathcal{F}_{d}^{(2)}(x)]+i\int dx_{1}\,[Q,R(W(x_{1});\mathcal{F}%
_{d}^{(1)}(x))] &&  \nonumber \\
+\frac{i^{2}}{2}\int dx_{1}dx_{2}\,[Q,R(W(x_{1})W(x_{2});F_{d}(x))] &=&0
\label{QF2}
\end{eqnarray}
The tree diagrams in the third term have the structure\footnote{%
Note that totally retarded products (\ref{retprod}) contain connected
diagrams only.} 
\begin{eqnarray}
&&R(W(x_{1})W(x_{2});F_{d}(x))|_{\mathrm{tree}}  \nonumber \\
&=&\sum \mathcal{D}_{1}^{\mathrm{ret}}(x-x_{1})\mathcal{D}_{2}^{\mathrm{ret}%
}(x_{1}-x_{2}):B_{1}(x_{1})B_{2}(x_{2})B_{3}(x_{2}):+(x_{1}\leftrightarrow
x_{2})  \label{tree2}
\end{eqnarray}
where $\mathcal{D}_{j}^{\mathrm{ret}}=\Delta ^{\mathrm{ret}},\,\partial
\Delta ^{\mathrm{ret}},\,(\partial \partial \Delta ^{\mathrm{ret}%
}+C_{j}\delta )\>(C_{j}$ are arbitrary constants) and with free fields $%
B_{k} $ (possibly with derivative), i.e. $B_{k}\in \{A^{\mu },\,\partial
A,\,\partial \partial A,(\partial )u,\,(\partial )\tilde{u},\,(\partial
)\phi ,\,(\partial )H\}$. All other diagrams in (\ref{QF2}) have less than
three legs at the vertex/vertices $\not=x$. Now let $y_{l}\in \mathbf{R}%
^{4},\>l=1,2,3$, with $(y_{l}-x)^{2}<0$ and let $\psi _{l},\>l=1,2,3$, be
arbitrary free fields. In the triple commutator of (\ref{QF2}) with $\psi
_{1}(y_{1}),\,\psi _{2}(y_{2})$ and $\psi _{3}(y_{3})$ only the terms (\ref
{tree2}) survive (the terms $\sim C_{1}\delta (x-x_{1})$ in (\ref{tree2}) do
not contribute either) 
\begin{eqnarray}
0 &=&\int dx_{1}dx_{2}\,[[[[Q,R(W(x_{1})W(x_{2});F_{d}(x))|_{\mathrm{tree}%
}],\psi _{1}(y_{1})],\psi _{2}(y_{2})],\psi _{3}(y_{3})]  \nonumber \\
&=&\int
dx_{1}dx_{2}\,[[[[Q,%
\{[F_{d}(x),T(W(x_{1})W(x_{2}))]-W(x_{1})[F_{d}(x),W(x_{2})]  \nonumber \\
&&-W(x_{2})[F_{d}(x),W(x_{1})]\}|_{\mathrm{tree}}],\psi _{1}(y_{1})],\psi
_{2}(y_{2})],\psi _{3}(y_{3})],  \label{Qtree2}
\end{eqnarray}
where we have inserted (\ref{retprod}), (\ref{antiT}) and the causal
factorization (\ref{2.4}) of the time ordered products due to $x\not\in
\{x_{1},x_{2}\}+\bar{V}^{-}$. Similarly to the step from (\ref{QF1'}) to (%
\ref{QF1''}) we conclude that $[Q,R(W(x_{1})W(x_{2});F_{d}(x))|_{\mathrm{tree%
}}]$ must be a sum of divergences (with respect to $x_{1}$ or $x_{2}$) of
local operators for $x_{1},x_{2}\in (x+\bar{V}^{-})\setminus \{x\}$. Now we
transform (\ref{Qtree2}) by means of identities of the kind 
\begin{eqnarray}
\lbrack Q,(M(x)N(y))|_{\mathrm{tree}}] &=&[Q,M(x)N(y)]|_{\mathrm{tree}} \\
&=&([Q,M(x)]N(y))|_{\mathrm{tree}}+(M(x)[Q,N(y)])|_{\mathrm{tree}},
\label{QC}
\end{eqnarray}
where $M$ and $N$ are arbitrary Wick polynomials. Then using (\ref{QF0}) and
(\ref{FQW}) we find the condition 
\begin{eqnarray}
\lbrack F_{d}(x),[Q,T(W(x_{1})W(x_{2}))]]|_{\mathrm{tree}}&-
&[Q,W(x_{1})][F_{d}(x),W(x_{2})]|_{\mathrm{tree}}  \nonumber \\
-[Q,W(x_{2})][F_{d}(x),W(x_{1})]|_{\mathrm{tree}} &=&\mathrm{div}%
_{x_{1}}+(x_{1}\leftrightarrow x_{2})  \label{QF2'}
\end{eqnarray}
for $x_{1},x_{2}\in (x+\bar{V}^{-})\setminus \{x\}$, where $\mathrm{div}_{y}$
means some divergence of a local operator with respect to $y$. Next we
specialize to the subregion $x_{1}^{0}>x_{2}^{0}$, where $%
T(W(x_{1})W(x_{2})) $ factorizes. By means of again (\ref{FQW}) we see that $%
W$ must satisfy 
\begin{equation}
\lbrack \lbrack F_{d}(x),W(x_{1})],[Q,W(x_{2})]]|_{\mathrm{tree}}=\mathrm{div%
}_{x_{1}}+\mathrm{div}_{x_{2}}\>\mathrm{for}\>x_{1},x_{2}\in (x+\bar{V}%
^{-})\setminus \{x\},\>x_{1}^{0}>x_{2}^{0}.  \label{QW2}
\end{equation}
Inserting the Ansatz for $W$ into this condition one finds that (\ref{QW2})
yields the same restrictions on the parameters in $W$ as the $Q$-divergence
condition to first order (\ref{2.29}): 
\begin{equation}
\lbrack Q,W]=\mathrm{divergence\>of\>a\>Wick\>polynomial}.  \label{QW}
\end{equation}
In other words we obtain the same results for the parameters in $W$ as in (%
\ref{2.62}), especially $f_{abc}=\epsilon _{abc}$. We recall that these
values of the parameters are not only necessary for (\ref{QW}), they are
also sufficient. Hence (\ref{FQW}) is also fulfilled. By inserting (\ref{QW}%
) into (\ref{QF2'}) we find the condition 
\begin{equation}
\lbrack F_{d}(x),[Q,T(W(x_{1})W(x_{2}))|_{\mathrm{tree}}]]=\mathrm{div}%
_{x_{1}}+(x_{1}\leftrightarrow x_{2})\quad \mathrm{for}\quad x_{1},x_{2}\in
(x+\bar{V}^{-})\setminus \{x\}.  \label{QWW}
\end{equation}
This is a necessary but not sufficient condition for $%
[Q,T(W(x_{1})W(x_{2}))|_{\mathrm{tree}}]=\mathrm{div}_{x_{1}}+(x_{1}%
\leftrightarrow x_{2})$. To get the full information of this latter
condition we need to go to third order.

- However, first we return to the "local terms" to first order (\ref{QF1lo}%
), because we shall need the validity of this equation. We explicitly
calculate the terms on the l.h.s. of (\ref{QF1''}), transform them into
divergence form (as far as possible) and obtain\footnote{%
By the experience of \cite{As} we know that all contributions to $P_d$ come
from the contraction of $F_d(x)$ with $\partial A(x_1)$ in the first term of 
$W(x_1)$ (\ref{ansatzW}).} 
\begin{equation}
P^{\mu\nu}_d=\epsilon_{bcd}[F^{\mu\nu}_b u_c-A^\mu_b\partial^\nu u_c+
A^\nu_b\partial^\mu u_c].  \label{P}
\end{equation}
Inserting this into (\ref{QF1lo}) we find that (\ref{QF1lo}) is fulfilled
iff the normalization 
\begin{equation}
C=-1  \label{C}
\end{equation}
is chosen and the first order coefficients in $\mathcal{F}^{\mu\nu}_{d\,%
\mathrm{int}}$ take the values 
\begin{eqnarray}
t^{(1)}_{dbc}=-\frac{1}{m_c}\epsilon_{bcd},\quad\tilde t^{(1)}=0, \quad
v^{(1)}=0, \quad\tilde v^{(1)}=0,\quad w^{(1)}=0,\quad\tilde w^{(1)}=0, 
\nonumber \\
w^{\prime (1)}=0,\quad \tilde x^{(1)}=0,\quad y^{(1)}=0,\quad\tilde
y^{(1)}=0,\quad y^{\prime (1)}=0, \quad \tilde z^{(1)}=0.  \label{F1}
\end{eqnarray}
$x^{(1)}$ and $z^{(1)}$ are still arbitrary. Note that (\ref{FQW}), (\ref{C}%
) and (\ref{F1}) are not only necessary for (\ref{QF1}), together they are
also sufficient.

- We now consider the condition (\ref{QF'}) to third order, i.e. the
analogous equation to (\ref{QF1}), (\ref{QF2}). By commuting this equation
with arbitrary free fields $\psi_1(y_1),\,\psi_2(y_2)$, $\psi_3(y_3),\,%
\psi_4(y_4), \>(y_l-x)^2<0\>\forall l$, we obtain 
\begin{eqnarray}
0=\frac{i^2}{2!}\int dx_2dx_3\,[[[[[Q,R(W(x_2)W(x_3);\mathcal{F}^{(1)}(x))]
,\psi_1(y_1)],\psi_2(y_2)],\psi_3(y_3)],\psi_4(y_4)]  \nonumber \\
+\frac{i^3}{3!}\int dx_1dx_2dx_3\,[[[[[Q,R(W(x_1)W(x_2)W(x_3);F(x))]
,\psi_1(y_1)],...\psi_4(y_4)].  \label{QF3}
\end{eqnarray}
Only tree diagrams of the following types contribute (we use the same
notations as in (\ref{tree2})): in the first term 
\begin{equation}
\sum \mathcal{D}^{\mathrm{ret}}_1(x-x_2)\mathcal{D}^{\mathrm{ret}}_2(x-x_3)
:B_0(x)B_1(x_2)B_2(x_2)B_3(x_3)B_4(x_3):+(x_2\leftrightarrow x_3)
\label{tree3}
\end{equation}
(where $B_0(x)$ is a free field or $\equiv 1$), in the second term 
\begin{equation}
\sum \mathcal{D}^{\mathrm{ret}}_3(x-x_1)\mathcal{D}^{\mathrm{ret}%
}_4(x_1-x_2) \mathcal{D}^{\mathrm{ret}%
}_5(x_1-x_3):B_5(x_2)B_6(x_2)B_7(x_3)B_8(x_3):+...  \label{tree4}
\end{equation}
and 
\begin{equation}
\sum \mathcal{D}^{\mathrm{ret}}_6(x-x_1)\mathcal{D}^{\mathrm{ret}%
}_7(x_1-x_2) \mathcal{D}^{\mathrm{ret}%
}_8(x_2-x_3):B_9(x_1)B_{10}(x_2)B_{11}(x_3)B_{12}(x_3):+...  \label{tree5}
\end{equation}
where the dots mean terms obtained by cyclic permutations. In the first term
in (\ref{QF3}) we take (\ref{retprod}) and the causal factorization basing
on $x\not\in (\{x_2,x_3\}+\bar V^-)$ into account. By means of (\ref{QW}) it
results 
\begin{eqnarray}
\frac{i^2}{2!}\int dx_2dx_3\,[[[[\{[[Q,\mathcal{F}%
^{(1)}(x)],T(W(x_2)W(x_3))] -W(x_2)[[Q,\mathcal{F}^{(1)}(x)],W(x_3)] 
\nonumber \\
-W(x_3)[[Q,\mathcal{F}^{(1)}(x)],W(x_2)]\}\vert_{\mathrm{tree}}
,\psi_1(y_1)]...,\psi_4(y_4)].  \label{QF31}
\end{eqnarray}
Additionally we have used that due to (\ref{tree3}) only the disconnected
diagram of $T(W(x_2)\break W(x_3))$ contributes, hence $[...[\mathcal{F}%
^{(1)}(x), [Q,T(W(x_2)W(x_3))]] ...,\psi_4(y_4)]=\mathrm{div}_{x_2}+\mathrm{%
div}_{x_3}$ by (\ref{QW}).

In the second term in (\ref{QF3}) the situation is more complicated. The
diagrams of the type (\ref{tree5}) obey the causal factorization basing on $%
x\not\in (\{x_{1},x_{2},x_{3}\}+\bar{V}^{-})$. But for the diagrams of the
type (\ref{tree4}) we only know $x\not\in (\{x_{2},x_{3}\}+\bar{V}^{-})$ (or
promulgated configurations). To fix the position of the third vertex we
consider two smooth functions $h_{1},\,h_{2}$ with 
\begin{eqnarray}
1 &=&h_{1}(y)+h_{2}(y)\>\forall y\in x+\bar{V}^{-},\quad h_{1}\in \mathcal{D}%
(\mathbf{R}^{4}),\quad x\in \mathrm{supp}\>h_{1},  \nonumber \\
\mathrm{supp}\>h_{2} &\cap &x+\bar{V}^{+}=\emptyset ,\quad
(y_{l}-z)^{2}<0\>\>\forall z\in \mathrm{supp}\>h_{1},\>l=1,2,3,4.  \label{h}
\end{eqnarray}
With that the second term in (\ref{QF3}) can be written in the form 
\begin{eqnarray}
&&\frac{i^{3}}{3!}\int
dx_{1}dx_{2}dx_{3}%
\,(h_{1}(x_{1})h_{2}(x_{2})h_{2}(x_{3})+h_{2}(x_{1})h_{1}(x_{2})h_{2}(x_{3})
+h_{2}(x_{1})h_{2}(x_{2})h_{1}(x_{3})
\nonumber \\
&&+h_{2}(x_{1})h_{2}(x_{2})h_{2}(x_{3}))[...[Q,R(W(x_{1})...;F(x))|_{\mathrm{%
tree}}]...,\psi _{4}(y_{4})].  \label{QF32}
\end{eqnarray}
In the terms with a factor $h_{1}(x_{l})$ the diagrams of the type (\ref
{tree4}) contribute only, and we may insert the causal factorization due to $%
\{x,x_{l}\}\cap (\{x_{j},x_{k}\}+\bar{V}^{-})=\emptyset $ (where $%
\{l,j,k\}=\{1,2,3\}$). So the contribution of $%
(h_{1}(x_{1})h_{2}(x_{2})h_{2}(x_{3})+h_{2}(x_{1})h_{1}(x_{2})h_{2}(x_{3})+
h_{2}(x_{1})h_{2}(x_{2})h_{1}(x_{3})) 
$ in (\ref{QF32}) reads 
\begin{eqnarray}
&&\frac{i^{3}}{2!}\int
dx_{1}dx_{2}dx_{3}\,h_{1}(x_{1})h_{2}(x_{2})h_{2}(x_{3})[...%
\{[[Q,R(W(x_{1});F(x))],T(W(x_{2})W(x_{3}))]  \nonumber \\
&&-W(x_{2})[[Q,R(W(x_{1});F(x))],W(x_{3})]  \nonumber \\
&&-W(x_{3})[[Q,R(W(x_{1});F(x))],W(x_{2})]\}|_{\mathrm{tree}}...,\psi
_{4}(y_{4})],  \label{h1}
\end{eqnarray}
where we have used $[Q,T(W(x_{2})W(x_{3}))]=\mathrm{div}_{x_{2}}+\mathrm{div}%
_{x_{3}}$, which holds true here for the same reason as in (\ref{QF31}).
>From (\ref{QF1}) and (\ref{RWF}) we know 
\begin{eqnarray}
\lbrack Q,\mathcal{F}^{(1)}(x)]+i\int
dx_{1}\,h_{1}(x_{1})[Q,R(W(x_{1});F(x))]=  \nonumber \\
-i\int dx_{1}\,h_{2}(x_{1})[Q,R(W(x_{1});F(x))]=-i\int
dx_{1}\,h_{2}(x_{1})\partial _{\tau }^{x_{1}}[F(x),W_{1}^{\tau }(x_{1})],
\label{QF1h}
\end{eqnarray}
where we have used $[Q,W]=\partial _{\tau }W_{1}^{\tau }$ (\ref{QW}). We now
insert (\ref{h1}), (\ref{QF32}) and (\ref{QF31}) into (\ref{QF3}). By means
of (\ref{QF1h}) we obtain 
\begin{eqnarray}
0 &=&\frac{i^{3}}{3!}\int
dx_{1}dx_{2}dx_{3}\,h_{2}(x_{1})h_{2}(x_{2})h_{2}(x_{3})[...%
\{[Q,R(W(x_{1})W(x_{2})W(x_{3});F(x))|_{\mathrm{tree}}]  \nonumber \\
&&-\partial _{\tau }^{x_{1}}([[F(x),W_{1}^{\tau
}(x_{1})],T(W(x_{2})W(x_{3}))]-W(x_{2})[[F(x),W_{1}^{\tau }(x_{1})],W(x_{3})]
\nonumber \\
&&-W(x_{3})[[F(x),W_{1}^{\tau }(x_{1})],W(x_{2})])|_{\mathrm{tree}}-\partial
^{x_{2}}(...)-\partial ^{x_{3}}(...)\}...,\psi _{4}(y_{4})],  \label{QF3'}
\end{eqnarray}
where $\partial ^{x_{2}}(...)$ and $\partial ^{x_{3}}(...)$ are obtained
from $\partial ^{x_{1}}(...)$ by cyclic permutation. The next step is
analogous to the step from (\ref{QF1'}) to (\ref{QF1''}), but more
complicated because one has to care about the cancelation of the boundary
terms. Taking the freedom in the choice of $h_{2}$ into account, (\ref{QF3'}%
) is equivalent to the existence of a local operator $R_{3}^{\tau
}(x_{1};x_{2},x_{3};x)$ with 
\begin{eqnarray}
\lbrack Q,R(W(x_{1})W(x_{2})W(x_{3});F(x))]|_{\mathrm{tree}} &=&\partial
_{\tau }^{x_{1}}R_{3}^{\tau }(x_{1};x_{2},x_{3};x)+\mathrm{%
cyclic\>permutations},  \nonumber \\
\forall x_{1},x_{2},x_{3} &\in &(x+\bar{V}^{-})\setminus \{x\},
\label{QF3''}
\end{eqnarray}
because the contribution from e.g. $\partial ^{x_{1}}(...)$ in (\ref{QF3'})
is equal to 
\begin{equation}
-\frac{i^{3}}{3!}\int dx_{1}dx_{2}dx_{3}\,(\partial _{\tau
}h_{2})(x_{1})h_{2}(x_{2})h_{2}(x_{3})[...R_{3}^{\tau
}(x_{1};x_{2},x_{3};x)...,\psi _{4}(y_{4})].  \label{dh2}
\end{equation}
Let us explain this latter statement. First note that due to (\ref{suppR})
and (\ref{h}) there is only a contribution in (\ref{dh2}) for $%
x_{1},x_{2},x_{3}\in (x+\bar{V}^{-})\setminus \{x\}\>\wedge \>x_{1}\not\in
(\{x_{2},x_{3}\}+\bar{V}^{-})$. In this region we have 
\begin{eqnarray}
R(W(x_{1})W(x_{2})W(x_{3});F(x)) &=&[[F(x),W(x_{1})],T(W(x_{2})W(x_{3}))] 
\nonumber \\
&&-W(x_{2})[[F(x),W(x_{1})],W(x_{3})]  \nonumber \\
&&-W(x_{3})[[F(x),W(x_{1})],W(x_{2})]  \label{R3}
\end{eqnarray}
due to causal factorization. Now we take into account that (in this region) $%
\partial _{\tau }^{x_{1}}R_{3}^{\tau }(x_{1};\break x_{2},x_{3};x)$ comes
from the $[Q,W(x_{1})]$-terms in $[Q,R(WWW;F)]$ (where $R(WWW;F)$ is given
by (\ref{R3})). However, by replacing $W(x_{1})$ by $\partial _{\tau
}W_{1}^{\tau }(x_{1})$ in (\ref{R3}) we obtain exactly the $\partial
^{x_{1}}(...)$-term in (\ref{QF3'}).

By means of causal factorization and (\ref{QF0}), (\ref{QW}), (\ref{QWW})
the condition (\ref{QF3''}) can be written in the form 
\begin{eqnarray}
&&%
\{[F(x),[Q,T(W(x_{1})W(x_{2})W(x_{3}))]]+[Q,T(W(x_{2})W(x_{3}))][W(x_{1}),F(x)]
\nonumber \\
&&+[Q,T(W(x_{1})W(x_{2}))][W(x_{3}),F(x)]+[Q,T(W(x_{1})W(x_{3}))][W(x_{2}),F(x)]\}|_{%
\mathrm{tree}}  \nonumber \\
&=&\mathrm{div}_{x_{1}}+\mathrm{cyclic\>permutations}  \label{QF3'''}
\end{eqnarray}
for $x_{1},x_{2},x_{3}\in (x+\bar{V}^{-})\setminus \{x\}$. Next we
specialize to the subregion $x_{1}\not\in (\{x_{2},x_{3}\}+\bar{V}^{-})$.
Again by causal factorization and (\ref{QWW}) we find that $%
T(W(x_{2})W(x_{3}))$ must fulfill 
\begin{equation}
\lbrack \lbrack F(x),W(x_{1})],[Q,T(W(x_{2})W(x_{3}))]]|_{\mathrm{tree}}=%
\mathrm{div}_{x_{1}}+\mathrm{div}_{x_{2}}+\mathrm{div}_{x_{3}}  \label{QWW'}
\end{equation}
for $x_{1},x_{2},x_{3}\in (x+\bar{V}^{-})\setminus \{x\}\>\wedge
\>x_{1}\not\in (\{x_{2},x_{3}\}+\bar{V}^{-})$. Inserting the explicit
expression for $W$ (with the values of the parameters obtained so far) one
sees that this condition is equivalent to 
\begin{equation}
\lbrack Q,T(W(x_{2})W(x_{3}))|_{\mathrm{tree}}]=\mathrm{div}%
_{x_{2}}+(x_{2}\leftrightarrow x_{3}).  \label{QWW''}
\end{equation}
As we mentioned in section 2 the latter requirement is satisfied iff the
masses agree $m\equiv m_{1}=m_{2}=m_{3}$ (\ref{2.47}), the parameters of the 
$H$-coupling take the values (\ref{2.48})\footnote{%
In the following calculations the undetermined sign $\kappa $ in (\ref{2.48}%
) is chosen to be $\kappa =1$.}, there is no term $\sim :H^4:$ in $W$ and if
the normalization constants in the tree normalization terms (\ref{2.49}) are
suitably chosen (up to $\lambda $ (\ref{2.50}) they are uniquely fixed by (%
\ref{QWW''})). So far there remain two free parameters in $W$ and $T(WW)|_{%
\mathrm{tree}}$: $s$ (\ref{ansatzW}) (cf. (\ref{2.48})) and $\lambda $ (\ref
{2.50}). In section 2 they have been determined by 
\begin{equation}
\lbrack Q,T(W(x_{1})W(x_{2})W(x_{3}))|_{\mathrm{tree}}]=\mathrm{div}_{x_{1}}+%
\mathrm{cyclic\>permutations}.  \label{QWWW}
\end{equation}
By inserting (\ref{QWW''}) into (\ref{QF3'''}) we find the weaker
requirement 
\begin{equation}
\lbrack F(x),[Q,T(W(x_{1})W(x_{2})W(x_{3}))|_{\mathrm{tree}}]]=\mathrm{div}%
_{x_{1}}+\mathrm{cyclic\>permutations}  \label{QWWW'}
\end{equation}
for $x_{1},x_{2},x_{3}\in (x+\bar{V}^{-})\setminus \{x\}$. But the latter
condition does not determined $s$ and $\lambda $. One needs to consider the
physicality condition (\ref{QF'}) to fourth order. There the values (\ref
{2.51}) for $s$ and $\lambda $ are obtained, as can be seen by an analogous
procedure. So we obtain exactly the same results for the parameters in $W$
and the tree normalizations in $T(WW)$ as in section 2. We recall that they
agree completely with the interaction Lagrangian obtained by the Higgs
mechanism.

- Now we have explained how to handle the essential difficulties which
appear in the determination of the parameters in $W$ (\ref{ansatzW}) and in $%
\mathcal{F}_{\mathrm{int}}$ (\ref{ansatzF}) by the physicality (\ref{QF'}).
We do not give further details. Instead we give a heuristic summary and the
remaining results. To $n$-th order the requirement (\ref{QF'}) reads 
\begin{equation}
0=[Q,\mathcal{F}_{d}^{(n)}(x)]+\sum_{l=1}^{n}\frac{i^{l}}{l!}\int
dx_{1}...dx_{l}\,[Q,R(W(x_{1})...W(x_{l});\mathcal{F}_{d}^{(n-l)}(x))]
\label{QFn}
\end{equation}
where $\mathcal{F}_{d}^{(0)}\equiv F_{d}$.

(I) In the "non-local terms" ($x_j\not= x$ for at least one $j$) we can
apply the causal factorization (\ref{2.4}) of the time ordered products.
Then the non-local terms in (\ref{QFn}) cancel due to 
\begin{equation}
[Q,T(W(x_1)...W(x_k))]=\,\mathrm{sum\>of\>divergences \> of \> local \>
operators},\quad 1\leq k\leq n,  \label{QT}
\end{equation}
and the physicality (\ref{QFn}) to lower orders $<n$ (see e.g. (\ref{QF1h}-%
\ref{QF3'})). We have seen that for the tree diagrams to lowest orders the
condition (\ref{QT}) is not only sufficient for the cancelation, it is also
necessary. But the lowest orders tree diagrams of (\ref{QT}) fix the
parameters in $W$ and $T(WW)\vert_{\mathrm{tree}}$, as was shown in \cite{As}%
.

For the cancelation of the "non-local terms" in (\ref{QFn}) it is not
important which kind of physical fields we require to exist, the properties
(i)-(vi) of $\mathcal{F}_{\mathrm{int}}$ (\ref{ansatzF}) are essentially not
needed. Hence we conjecture that already the existence of \textit{any}
non-trivial observable fixes the interaction.

(II) The cancelation of the remaining terms in (\ref{QFn}), which are
"local" (i.e. their support is the total diagonal $x_j=x,\> \forall j$),
requires a suitable normalization of the time ordered products $%
T(W(x_1)...W(x_l);\mathcal{F}^{(n-l)}_d(x))$ (see e.g. (\ref{C})) and a
suitable choice of the parameters in $\mathcal{F}^{(n-l)}_d,\>l=0,1,...,n-1$.%
\footnote{%
Mainly parameters in $\mathcal{F}^{(n)}_d$ are determined; most of the
parameters in $\mathcal{F}^{(n-l)}_d,\>1\leq l\leq n-1$, have already been
fixed in earlier steps of the induction.} In this way the latter parameters
are \textit{uniquely} determined (at least to lowest orders).

So the cancelation of the "local terms" to second order requires 
\begin{eqnarray}
t^{(2)}=0,\quad\tilde t^{(2)}=0,\quad v^{(2)}_{dabc}=-\frac{1}{2m^2} [%
(\delta_{da}-\frac{m}{2}x^{(1)}_{da})\delta_{bc}-\frac{1}{2}(\delta_{dc}
\delta_{ba}+\delta_{db}\delta_{ca})],  \nonumber \\
\tilde v^{(2)}=0,\quad w^{(2)}=0,\quad\tilde w^{(2)}=0,\quad w^{\prime
(2)}=0,\quad\tilde x^{(2)}=0,  \nonumber \\
y^{(2)}_{dbc}=\frac{1}{2m^2}\epsilon_{dbc}-\frac{1}{m}x^{(1)}_{da}
\epsilon_{abc},\quad\tilde y^{(2)}=0, \quad y^{\prime (2)}=0,\quad \tilde
z^{(2)}=0,\quad z^{(1)}=0.  \label{F2}
\end{eqnarray}
The very last equation results from the terms $\sim g^2:F(x)u(x)H^2(x):$. In
this case the cancelation is of a special kind as explained below. The
other equations are obtained analogously to (\ref{QF1lo}), (\ref{F1}). They
come from bilinear and trilinear terms. $x^{(1)},\,x^{(2)}$ and $z^{(2)}$
are still arbitrary.

To third order the terms $\sim g^3:F(x)u(x)\phi(x)\phi(x):$, ($\sim
g^3:F(x)u(x)H^2(x):$ respectively) cancel iff 
\begin{equation}
x^{(1)}_{da}=\frac{1}{m}\delta_{da},\quad\quad (z^{(2)}_{da}=\frac{1}{4m^2}%
\delta_{da}\>\>\mathrm{respectively}).  \label{F3}
\end{equation}
Similarly one finds to fourth order that $x^{(2)}$ must vanish.

Summing up (\ref{F1}), (\ref{F2}) and (\ref{F3}) we obtain the result (\ref
{resultF}). But we have not checked that the cancelation of the ''local
terms'' in (\ref{QFn}) can be achieved for loop diagrams and for tree
diagrams to higher orders. For the tree diagrams with more than three
external legs (e.g. the very last equation in (\ref{F2}) and the terms
considered in (\ref{F3})) there is no contribution from $\mathcal{F}%
_{d}^{(n)}$ in (\ref{QFn}) due to our Ansatz (\ref{ansatzF}). Hence the
parameters in $\mathcal{F}_{d}^{(n-l)},\>l\geq 1$ are available only
(besides normalization constants). However, most of the latter parameters
have already been fixed (or restricted) by (\ref{QFn}) to lower orders. (For
example to achieve the cancelation of the terms considered in (\ref{F3}),
it is important that $x^{(1)}$ ($z^{(2)}$ resp.) has not been fixed
previously.) So we have shown the existence of observables $\mathcal{F}_{d\,%
\mathrm{int}}^{\mu \nu },\>d=1,2,3,$ only partially. But there is a strong
hint that this holds true for all terms to all orders from the
identification (given in the main text) of our physical fields $\mathcal{F}%
_{d\,\mathrm{int}}^{\mu \nu }$ (\ref{resultF}) as gauge invariant fields in
the framework of spontaneous symmetry breaking of the $SU(2)$ gauge symmetry.

\end{document}